\documentclass[fleqn,usenatbib]{mnras}

\usepackage{newtxtext,newtxmath}

\usepackage[T1]{fontenc}

\DeclareRobustCommand{\VAN}[3]{#2}
\let\VANthebibliography\thebibliography
\def\thebibliography{\DeclareRobustCommand{\VAN}[3]{##3}\VANthebibliography}

\usepackage{graphicx}	
\usepackage{amsmath}	

\usepackage{xcolor}
\usepackage{overpic}
\usepackage{subfigure}
\usepackage{booktabs}
\usepackage{multirow}
\usepackage{caption}
\usepackage[flushleft]{threeparttable}

\newcommand{\ms}{m\,s$^{-1}$}

\title[Luminosity and velocity oscillation amplitudes]{The relationship between photometric and spectroscopic oscillation amplitudes from 3D stellar atmosphere simulations}

\author[Y. Zhou et al.]{%
Yixiao Zhou,$^{1}$
Thomas Nordlander,$^{1,2}$ 
Luca Casagrande,$^{1,2}$ 
Meridith Joyce,$^{1,2}$
\newauthor
Yaguang Li,$^{3,4}$
Anish M. Amarsi,$^{5}$
Henrique Reggiani$^{6}$
and Martin Asplund$^{7}$
\\
$^{1}$Research School of Astronomy and Astrophysics, Australian National University, Canberra, ACT 2611, Australia\\
$^{2}$ARC Centre of Excellence for All Sky Astrophysics in 3 Dimensions (ASTRO 3D), Australia\\
$^{3}$Sydney Institute for Astronomy (SIfA), School of Physics, University of Sydney, NSW 2006, Australia\\
$^{4}$Stellar Astrophysics Centre, Department of Physics and Astronomy, Aarhus University, Ny Munkegade 120, DK-8000 Aarhus C, Denmark\\
$^{5}$Theoretical Astrophysics, Department of Physics and Astronomy, Uppsala University, Box 516, SE-751 20 Uppsala, Sweden \\
$^{6}$Department of Physics and Astronomy, Johns Hopkins University, 3400 N Charles St., Baltimore, MD 21218, USA\\
$^{7}$Max Planck Institute for Astrophysics, Karl-Schwarzschild-Str. 1, D-85741 Garching, Germany
}

\date{Accepted XXX. Received YYY; in original form ZZZ}

\pubyear{2021}

\begin{document}
\label{firstpage}
\pagerange{\pageref{firstpage}--\pageref{lastpage}}
\maketitle

\begin{abstract}
We establish a quantitative relationship between photometric and spectroscopic detections of solar-like oscillations using \textit{ab initio}, three-dimensional (3D), hydrodynamical numerical simulations of stellar atmospheres. We present a theoretical derivation as proof of concept for our method. We perform realistic spectral line formation calculations to quantify the ratio between luminosity and radial velocity amplitude for two case studies: the Sun and the red giant $\epsilon$ Tau. Luminosity amplitudes are computed based on the bolometric flux predicted by 3D simulations with granulation background modelled the same way as asteroseismic observations. Radial velocity amplitudes are determined from the wavelength shift of synthesized spectral lines with methods closely resembling those used in BiSON and SONG observations. 
Consequently, the theoretical luminosity to radial velocity amplitude ratios are directly comparable with corresponding observations. 
For the Sun, we predict theoretical ratios of 21.0 and 23.7 ppm/[\ms] from BiSON and SONG respectively, in good agreement with observations 19.1 and 21.6 ppm/[\ms].  For $\epsilon$ Tau, we predict \textit{K2} and SONG ratios of 48.4 ppm/[\ms], again in good agreement with observations 42.2 ppm/[\ms], and much improved over the result from conventional empirical scaling relations which gives 23.2 ppm/[\ms]. This study thus opens the path towards a quantitative understanding of solar-like oscillations, via detailed modelling of 3D stellar atmospheres.
\end{abstract}




\section{Introduction}

  Solar-like oscillations can be observed via photometry and spectroscopy. The photometric method allows us to detect stellar oscillations by measuring variations in the brightness of stars, whereas the spectroscopic method exploits the Doppler shifts of spectral lines to detect stellar oscillations. The spectroscopic method (also called the radial velocity method) is employed by ground-based telescopes such as Birmingham Solar Oscillations Network (BiSON, \citealt{1996SoPh..168....1C}) and Stellar Oscillations Network Group (SONG, \citealt{2006MmSAI..77..458G}). These instrumments have laid the groundwork for helioseismoloy and asteroseismology through their detailed observations of solar oscillations and their detection of the first solar-like oscillating stars \citep{1979Natur.282..591C,1991ApJ...368..599B,2001ApJ...549L.105B,2003AJ....126.1483K}. 
  
   Built upon these pioneering works, the field of asteroseismology has thrived in the last decade thanks to the CoRoT \citep{2008Sci...322..558M}, \textit{Kepler} \citep{2010Sci...327..977B} and TESS \citep{2015JATIS...1a4003R} missions that detect stellar oscillations by measuring variations in stellar luminosity. The high-quality, long time-series, extensive photometric data provided by these space-based telescopes enable accurate determination of oscillation frequencies and amplitudes for thousands of solar-like stars, thus ushering in the era of ensemble asteroseismology. However, in the low-frequency regime the photometric method is complicated by signals due to stellar atmospheric convection (stellar granulation), which impedes the characterisation of low-frequency oscillations. This difficulty can be avoided by observing the star using the radial velocity method, as stellar granulation noise is significantly less pronounced in velocity signals. Moreover, the radial velocity method has demonstrated great potential for measuring oscillations in cool dwarf stars (e.g.~\citealt{2005ApJ...635.1281K}), which are important in exoplanet science but difficult to detect with space-photometry due to their low intrinsic luminosity and small oscillation amplitude \citep{2019arXiv190308188H}. Nonetheless, ground-based spectroscopy is limited by target brightness and the Earth's atmosphere.
  
  It follows that the photometric and spectroscopic methods of measuring stellar oscillations are highly complementary and that combining the two methods will yield extra information that can further constrain the properties of stars. Recently, solar-like oscillations in several stars, such as Procyon A and $\epsilon$ Tauri (hereafter $\epsilon$ Tau), have been observed in both photometry and spectroscopy \citep{2011ApJ...731...94H, 2019A&A...622A.190A}. With the commencement of the TESS mission and SONG observations, many stars will soon have both luminosity and radial velocity data available. Therefore, investigating the relationship between luminosity and radial velocity amplitude is of increasing importance.
 
  This topic was first explored in the pioneering study of solar-like oscillations by \citet{1995A&A...293...87K}, who proposed a quantitative relationship between luminosity and radial velocity oscillation amplitudes for solar-like stars by scaling from the Sun. The \citet{1995A&A...293...87K} amplitude ratio scaling relation has been the industry standard in asteroseismology until now, providing valuable guidance for many years. However, their relationship is based on empirical arguments, and it is unable to reproduce the observed amplitude ratio for some stars \citep{2011ApJ...731...94H,2019A&A...622A.190A}. It is therefore prudent and timely to refine the relationship between luminosity and radial velocity based on detailed stellar modelling. As a first attempt to solve this problem from a modelling perspective, \citet{1999A&A...351..582H} and \citet{2010Ap&SS.328..237H} computed the theoretical ratio between luminosity and velocity amplitudes. The calculations are based on their one-dimensional, non-local, time dependent convection model for the Sun \citep{1999A&A...351..582H} and the scaled VAL-C atmosphere for Procyon A \citep{1981ApJS...45..635V,2010Ap&SS.328..237H}. Their amplitude ratio results are in reasonable agreement with observations. Nevertheless, it is worth noting that the predicted amplitude ratio depends on at which atmospheric height the velocity amplitude is evaluated (see Fig.~1 and 2 of \citealt{2010Ap&SS.328..237H}).

  In this paper, we investigate the relationship between photometric and spectroscopic measurements of stellar oscillations. We quantify the amplitude ratio in an essentially parameter-free manner, by carrying out detailed \textit{ab initio} three-dimensional (3D) hydrodynamical simulations of stellar surface convection. We base our analysis on realistic synthetic spectra, calculated using 3D radiative transfer and taking into account departures from local thermodynamic equilibrium (LTE) where necessary.

\section{Observational Data} \label{sec:obs}

  In this pilot study, we focus on the Sun and on the G-type red giant star $\epsilon$ Tau (HD 28305). As a bright star residing in the nearest open Cluster, Hyades, and known exoplanet host, $\epsilon$ Tau is of great interest to stellar physics for a variety of reasons \citep{2007ApJ...661..527S}. 
  
  We adopt the stellar parameters provided in \citet{2019A&A...622A.190A}: $T_{\rm eff} = 4976$ K, $\log g = 2.67$ dex, $\rm [Fe/H] = 0.15$ dex, as reference values. This effective temperature was determined via the bolometric flux measured by \citet{2018AJ....155...30B} and the angular diameter measured interferometrically from the CHARA array \citep{2019A&A...622A.190A}. The surface gravity was determined from the observed frequency of maximum power, $\nu_{\max}$, for this star \citep{2017MNRAS.472.4110S,2019A&A...622A.190A} through the $\nu_{\max}$ scaling relation \citep{1991ApJ...368..599B,1995A&A...293...87K}. Moreover, detailed asteroseismic observations for $\epsilon$ Tau using both $\textit{K2}$ (the successor of \textit{Kepler}; \citealt{2014PASP..126..398H}) and SONG yield individual oscillation frequencies for more than 20 modes as well as the amplitude ratio between $\textit{K2}$ and SONG, which makes $\epsilon$ Tau an ideal target to investigate in this work. Analogous parameters for the Sun are, of course, known to the highest degrees of precision and accuracy of any star. Observational parameters are included in Table \ref{tb:simu_info}.

\section{Three-dimensional stellar atmosphere models} \label{sec:3Dmodel}

\begin{table}
\centering
\caption{Fundamental parameters and basic information about the simulation of the Sun and $\epsilon$ Tau. Reference values are adopted from \citet{2016AJ....152...41P} and \citet{2019A&A...622A.190A}, respectively. We note that the effective temperature fluctuates over time in 3D models, therefore both mean effective temperature and its standard deviation are given. Also, both minimum and maximum vertical grid spacing are provided, as mesh points are not uniformly distributed vertically.
\label{tb:simu_info}}
{\begin{tabular*}{\columnwidth}{@{\extracolsep{\fill}}cccc}
\toprule[2pt]
  \multicolumn{2}{c}{} & Sun & $\epsilon$ Tau
  \\
\midrule[1pt]
  \multirow{2}{*}{$T_{\rm eff}$ (K)} & Reference & $5772.0 \pm 0.8$ & $4976 \pm 63$
  \\ 
   & Modelling & $5773 \pm 16$ & $4979 \pm 18$
  \\ 
  \multirow{2}{*}{$\log g$ (cgs)} & Reference & 4.438 & 2.67 
  \\  
   & Modelling & 4.438 & 2.67
  \\ 
  \multirow{2}{*}{[Fe/H] (dex)} & Reference & 0.00 & $0.15 \pm 0.02$
  \\
   & Modelling & 0.00 & 0.00
  \\
  
  \multicolumn{2}{c}{Numerical resolution} & $240^3$ & $240^3$
  \\
  \multicolumn{2}{c}{Time duration (hour)} & 24 & 1205.8
  \\
  \multicolumn{2}{c}{Sampling interval (s)} & 30 & 1447
  \\
  \multicolumn{2}{c}{Vertical size (Mm)} & 3.6 & 250
  \\
  \multicolumn{2}{c}{Vertical grid spacing (km)} & 7--33 & 562--2650
  \\
  \multicolumn{2}{c}{Horizontal grid spacing (km)} & 25 & 2165
  \\
\bottomrule[2pt]
\end{tabular*}}
\end{table}

\begin{figure*}
\subfigure[]{
\begin{overpic}[width=0.48\textwidth]{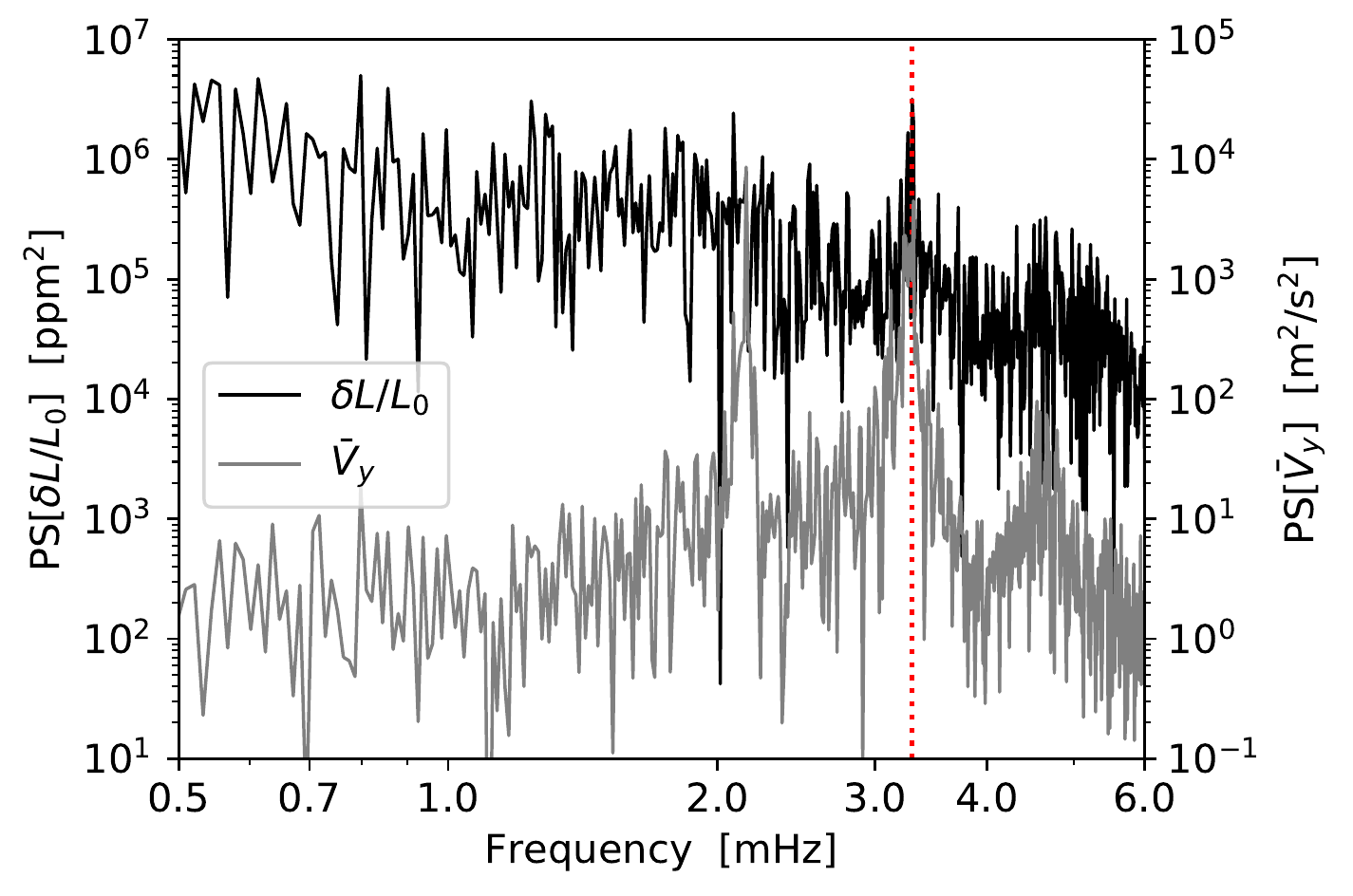}
\end{overpic}
\label{fig:sunPS}
}
\subfigure[]{
\begin{overpic}[width=0.48\textwidth]{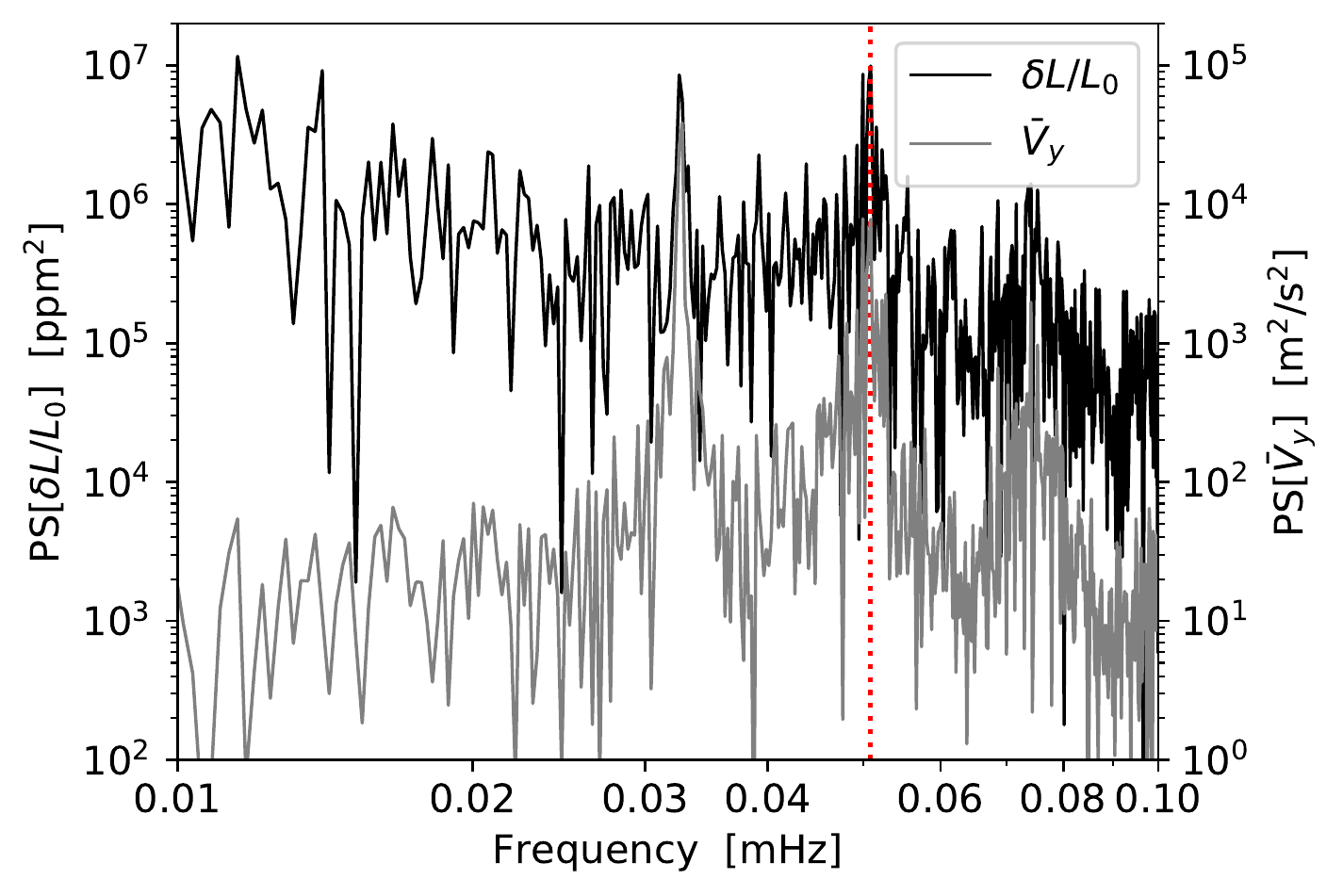}
\end{overpic}
\label{fig:t50g27m00PS}
}
\caption{\ref{fig:sunPS}: The power spectrum of luminosity amplitude (black line) calculated from the 3D solar atmosphere model is plotted together with the power spectrum of the horizontally averaged vertical velocity amplitude around the photosphere (grey line). Three simulation modes are clearly seen in the velocity spectrum, whereas only the simulation mode with frequency slightly greater than 3 mHz is recognizable in the luminosity spectrum. Red dotted vertical line indicates the frequency of the dominant simulation mode. \ref{fig:t50g27m00PS}: Power spectra calculated from the 3D atmosphere model for $\epsilon$ Tau. Three simulation modes are visible in both the velocity and luminosity spectra. 
}
\label{fig:PS}
\end{figure*}

  In this section, we introduce the 3D hydrodynamic stellar atmosphere models that are the basis of our analysis. All 3D models are computed with a customized version of the \textsc{Stagger} code \citep{1995...Staggercodepaper,2018MNRAS.475.3369C}, a radiative-magnetohydrodynamic code that solves the time-dependent equations of mass, momentum and energy conservation, as well as the magnetic-field induction equation and the radiative transfer equation on a 3D staggered Eulerian mesh. The stellar models in the present study have been constructed without magnetic fields. 
  All scalars are evaluated at cell centres, whereas vectors, such as velocity, are staggered at the centres of cell faces in order to improve numerical accuracy. The code incorporates realistic microphysics and a detailed radiative transfer scheme. An updated version of the \citet{1988ApJ...331..815M} equation of state \citep{2013ApJ...769...18T} is adopted, which accounts for all ionization stages of the 17 most abundant elements in the Sun plus the H$_2$ molecule. A comprehensive collection of relevant continuous absorption and scattering sources is included as described in \citet{2010A&A...517A..49H}. The pre-computed, sampled line opacities are taken from the MARCS model atmosphere package \citep{2008A&A...486..951G}.
  Radiative energy transport is modelled by solving the equation of radiative transfer at every time step of the simulation for all mesh points above a certain Rosseland mean optical depth ($\tau_{\rm Ross} \leq 500$ throughout this work) under the assumption of LTE. The frequency dependence of the radiative transfer equation is approximated via the opacity binning method \citep{1982A&A...107....1N,2018MNRAS.475.3369C}, in which 12 opacity bins are divided based on wavelength and strength of opacities. Consequently, the integration over wavelength reduces to the summation over 12 selected bins. The spatial dependence of the radiative transfer equation is represented by solving along a set of inclined rays in space. Nine directions -- one vertical and eight inclined directions representing combinations of two polar and four azimuthal angles -- are considered for all models presented in this work. The integration over polar angle is carried out using the Gauss-Radau quadrature scheme. The thus evaluated radiative heating rates can be used to calculate the surface flux (i.e., the emergent radiative flux at the top boundary of simulation domain) and subsequently the effective temperature via the Stefan-Boltzmann law.
  
  The basic configurations of our 3D models are summarised in Table \ref{tb:simu_info}. For both the Sun and the red giant $\epsilon$ Tau, the \textsc{Stagger} model atmospheres are constructed based on the reference effective temperatures and surface gravities (Table \ref{tb:simu_info}). The \citet{2009ARA&A..47..481A} solar chemical composition is adopted in both cases. Though we do not expect stellar metallicity to introduce any significant differences for the purposes of this study, we intend to consider the effects of stellar metallicity in detail in a later investigation.  
  
  Spatially, the simulation domain is discretized in a box located around the stellar photosphere. Horizontally, the simulation domain is a square with $240 \times 240$ evenly distributed mesh points. The horizontal size of the box is large enough to enclose at least ten granules at any time of the simulation \citep{2013A&A...557A..26M}. There are 240 mesh points in the vertical direction covering roughly the outer 1\% of the stellar radius, extending from the upper part of the surface convection zone, including the entire optical surface, and reaching the lower part of the chromosphere; we note that the outer-most layers are likely the least realistic given our neglect of magnetic fields in these simulations. Because the vertical scale of the simulation is very small compared to the total stellar radius, the spherical effects are negligible and gravitational acceleration can be regarded as a constant (i.e.~the surface gravity). Mesh points are not evenly distributed vertically: the highest numerical resolution is applied around the optical surface to resolve the transition between the optically thick and thin regimes. 
  Furthermore, in the case of $\epsilon$ Tau, a separate vertical mesh structure is employed for the radiative transfer calculation in order to resolve the extremely steep temperature and opacity gradients near the optical surface of red giants adequately (see e.g., Fig.~3 of \citealt{2018MNRAS.475.3369C}). 
  Adaptive mesh refinement was used when constructing the vertical radiative mesh; the radiative mesh of each vertical sub-domain within the simulation domain is arranged based on the distribution of Rosseland optical depth in this sub-domain, resulting in highest numerical resolution near the photosphere (see Fig.~6 in \citealt{2018MNRAS.475.3369C} for an illustration). At each simulation time step, radiative transfer calculations are performed on the radiative mesh and then interpolated back to the aforementioned hydrodynamical mesh. We refer the reader to Section 2.7  in \citet{2018MNRAS.475.3369C} for a detailed introduction to this technique.
  
  Boundaries are periodic in the horizontal direction while open in the vertical \citep{2018MNRAS.475.3369C}. At the bottom boundary, outgoing flows (vertical velocities towards stellar centre) are free to carry their entropy fluctuations out of the simulation domain, whereas incoming flows have invariant entropy and thermal (gas plus radiation) pressure.
  Temporally, the duration of the simulation is one day for the Sun, and about 50 days for $\epsilon$ Tau. Simulation data is stored every 30 seconds in the solar simulation while every 1447 seconds for the red giant case. A long stellar time coverage like this is necessary for an accurate analysis of stellar oscillations. 
  
  Sound waves and the resulting $p$-modes are natural phenomena in surface convection simulations, which can be directly identified by looking at the power spectrum of the vertical velocity of the simulations. Because $p$-mode oscillations in the simulation domain periodically shift the optical surface up and down, causing coherent changes in surface temperature, simulation modes can also be identified indirectly from the power spectrum of the bolometric flux variation. The relative variation of the bolometric flux, in parts per million (ppm), is defined as 
\begin{equation} \label{eq:dL_def}
\frac{\delta F_{\rm bol}}{F_{\rm bol,0}} = 
\frac{F_{\rm bol} - F_{\rm bol,0}}{F_{\rm bol,0}} \times 10^6 = 
\frac{\delta L}{L_0}.
\end{equation}
This is essentially equivalent to the relative variation in luminosity $\delta L / L_0$ (in ppm) because oscillations hardly change the total stellar radius. The subscript ``0'' indicates time-averaged quantities, i.e.~the equilibrium state. 

  The power spectra (PS) of the vertical velocity variation and the relative luminosity variation (luminosity spectrum for short hereinafter) are computed via
\begin{equation}
\begin{aligned}
{\rm PS}[f](\omega) &= \frac{4}{N^2}
\left\vert \sum\limits_{s=0}^{N-1} f(t_s) e^{i\omega s \Delta t} \right\vert^2
\\
\quad \omega & = \frac{2\pi}{N\Delta t} \left(1,2,..., \frac{N}{2} - 1 \right)
\\
f &= \bar{V}_y \quad \text{or} \quad \frac{\delta L}{L_0}.
\end{aligned}
\end{equation}
Here, $\bar{V}_y$ is the horizontally averaged vertical velocity, and the terms $t$, $\Delta t$ and $\omega$ are time, time interval between two consecutive snapshots and angular frequency, respectively. The symbol $s$ denotes individual simulation snapshots, and $N$ is the total number of snapshots. 

  Fig.~\ref{fig:PS} shows the results for the solar and the red giant simulations. Three radial simulation modes with frequencies of approximately 2.1, 3.3, and 4.7 mHz are seen in the vertical velocity spectrum of the solar simulation. 
These are the fundamental, first overtone, and second overtone radial modes in the simulation box, respectively. Among the three, only the intermediate-frequency, first overtone simulation mode is clearly recognizable from the luminosity spectrum. The reason is that at low frequencies, the granulation signal is relatively strong, causing the signature of the low-frequency simulation mode to be overwhelmed by ``convective noise.'' On the other hand, the amplitude of the high-frequency simulation mode is too small to be clearly identified in the luminosity spectrum (black line in Fig.~\ref{fig:sunPS}). We therefore refer to the first overtone radial mode as the dominant simulation mode, as it is the only one that is identifiable in both the vertical velocity and luminosity spectra in the solar case. The situation for the $\epsilon$ Tau simulation is different. The three simulation modes are visible in both the velocity and luminosity spectra owing to their large amplitude. 
We note that for both the Sun and $\epsilon$ Tau, the duration of the simulation is long enough to cover at least 200 periods of the dominant simulation mode. Likewise, the sampling interval is short enough in both cases such that at least 10 snapshots are stored within one pulsation cycle of the dominant simulation mode. These two factors together ensure that the dominant simulation mode is well resolved in the frequency domain. The exact frequency of the dominant simulation mode is important for the analysis below (Sect.~\ref{sec:bol} and \ref{sec:RV}). It is determined by looking for the local maximum $\bar{V}_y$ for all vertical layers in the simulation domain, which is similar to the method used in \citet{2019A&A...625A..20B}. The exact frequency values are 3.299 mHz and 0.051 mHz for the solar and $\epsilon$ Tau simulation respectively, which are also highlighted in red dotted lines in Fig.~\ref{fig:PS}.
  
  It is worth noting that the amplitude of simulation mode is on the order of 100\,\ms. This is much greater than the observed amplitude of radial $p$-modes as measured in the solar flux spectrum, which is around 0.2\,\ms. This difference in amplitude between the simulation mode and the observed stellar $p$-mode was explained in detail in \citet{2019A&A...625A..20B} and \citet{2019ApJ...880...13Z}. In short, the discrepancy emerges from the difference between the volume of the simulation box and the volume of the real star. Stellar $p$-modes propagate throughout the entire stellar surface and interior, whereas the simulation modes are confined to the simulation box whose horizontal and vertical extents are significantly smaller than the dimensions of a star. Therefore, the luminosity or velocity amplitudes from 3D atmosphere simulations are not directly comparable to the corresponding asteroseismic observations. A natural question is then whether realistic ratio between luminosity and velocity amplitude can be predicted from our simulations? We address this question in detail in the subsequent section.

\section{Proof of concept} \label{sec:Proof}

  We demonstrate in this section that, in principle, 3D surface convection simulations are able to reliably predict the relationship between the luminosity and velocity amplitudes (the amplitude ratio) despite their individual values not being comparable with observations. We begin with the relative luminosity variation defined in Eq.~\eqref{eq:dL_def}. Assuming the source function in the radiative transfer equation $S_{\nu}$ ($\nu$ is the radiation frequency) is a linear function of optical depth $\tau_{\nu}$ (i.e. equivalent to the Eddington-Barbier approximation), the surface flux at a given frequency is given by:
\begin{equation}
F_{\nu}(\tau_{\nu}=0) = \pi S_{\nu}(\tau_{\nu}=2/3).
\end{equation}
Further assuming LTE gives
\begin{equation}
F_{\nu}(\tau_{\nu}=0) = \pi B_{\nu}(\tau_{\nu}=2/3) = \pi \frac{2h \nu^3}{c^2} 
\frac{1}{\exp \left[ \frac{h\nu}{k_B T(\tau_{\nu}=2/3)}\right] - 1}.
\end{equation}
Here, $B_{\nu}$ is the Planck function, $c$, $h$ and $k_B$ are speed of light, Planck constant and Boltzmann constant, respectively. The term $T(\tau_{\nu}=2/3)$ is the temperature at optical depth $\tau_{\nu}=2/3$. Because at different frequencies, the $\tau_{\nu}=2/3$ layer corresponds to different locations in the stellar atmosphere due to opacity variations, $T(\tau_{\nu}=2/3)$ depends on frequency in general. In the case of a grey atmosphere where optical depth has no frequency dependence, the integration of $F_{\nu}$ over frequency gives the Stefan-Boltzmann law. The bolometric flux is hence
\begin{equation} \label{eq:SBlaw}
F_{\rm bol} = \pi \int_0^{\infty} B_{\nu}(\tau = 2/3) \: d\nu 
= \sigma T^4(\tau = 2/3),
\end{equation}
where $\sigma$ is the Stefan-Boltzmann constant. Combining Eqs.~\eqref{eq:dL_def} and \eqref{eq:SBlaw} yields
\begin{equation} \label{eq:dL-T}
\frac{\delta L}{L_0} = \frac{4 \delta T(\tau=2/3)}{T_0(\tau=2/3)} \times 10^6,
\end{equation}
where $\delta T$ denotes temperature fluctuation at constant optical depth. From Eq.~\eqref{eq:dL-T} we can then recognise that the luminosity variation essentially captures the fluctuation in temperature at the optical surface. For solar-type stars without strong stellar activity, such fluctuation is due primarily to surface convection and secondarily due to acoustic oscillations. The contribution due to surface convection will be separated from the acoustic oscillations in Sect.~\ref{sec:Abol}.

  Next we connect fluid velocity $V$ with the fluctuations of thermodynamical quantities. Following the discussion in \citet{2010aste.book.....A} (see Chapter 3.1.4), we assume $V$ is caused solely by sound waves and is small compared to the sound speed. It is worth noting that convective velocities are non-negligible in stellar convection zones; their magnitude can even be comparable to the local sound speed in the near-surface region. Nevertheless, convective velocities are effectively regarded as ``equilibrium state,'' since oscillation is the focus here. Under this assumption, density $\rho$, pressure $P$ and temperature $T$ can be written as $f = f_0 + f^{\prime}$, where $f^{\prime}$ is the small Eulerian perturbation\footnote{Perturbations at constant geometric depth (or radius)} (second and higher order terms are ignored). After further assuming that the medium is spatially homogeneous, all derivatives of equilibrium quantities vanish. The fluid continuity equation
\begin{equation}
\frac{\partial \rho}{\partial t} + \nabla \cdot (\rho \vec{V}) = 0
\end{equation}
then becomes
\begin{equation} \label{eq:ce}
\frac{\partial \rho^{\prime}}{\partial t} + \rho_0 \nabla \cdot \vec{V} = 0,
\end{equation}
while the equation of motion 
\begin{equation}
\rho \frac{\partial \vec{V}}{\partial t} + \rho\vec{V} \cdot \nabla\vec{V} 
= -\nabla P + \rho \vec{g}
\end{equation}
becomes
\begin{equation}
\rho_0 \frac{\partial \vec{V}}{\partial t} = 
-\nabla P^{\prime} + \rho_0 \vec{g}^{\prime},
\end{equation}
where $g$ is gravitational acceleration. If we ignore the perturbation to gravitational acceleration $\vec{g}^{\prime}$ (i.e.~Cowling approximation), this simplifies to
\begin{equation} \label{eq:eom}
\rho_0 \frac{\partial \vec{V}}{\partial t} + \nabla P^{\prime} = 0.
\end{equation}
Now taking the time derivative of Eq.~\eqref{eq:ce} and making use of Eq.~\eqref{eq:eom}, we have
\begin{equation} \label{eq:waveeq_pre}
\frac{\partial^2 \rho^{\prime}}{\partial t^2} - \nabla^2 P^{\prime} = 0.
\end{equation}
In the case of adiabatic oscillation, pressure and density fluctuations are connected by
\begin{equation} \label{eq:energyeq_ad}
P^{\prime} = c_{s,0}^2 \rho^{\prime},
\end{equation}
where $c_s = \sqrt{(\partial P / \partial\rho)_{\rm ad}}$ is the adiabatic sound speed. Substituting Eq.~\eqref{eq:energyeq_ad} into Eq.~\eqref{eq:waveeq_pre} gives the wave equation (\citealt{2010aste.book.....A} Eq.~3.51):
\begin{equation}
\frac{\partial^2 \rho^{\prime}}{\partial t^2} - c_{s,0}^2\nabla^2 \rho^{\prime} = 0.
\end{equation}

If we now consider a pure radial sound wave in which all quantities depend only on the $y$-coordinate, then density and pressure fluctuation can be written as
\begin{equation} \label{eq:rhoP}
\begin{aligned}
\rho^{\prime} &= a \cos(ky - \omega t),
\\
P^{\prime} &= c_{s,0}^2 a \cos(ky - \omega t),
\end{aligned}
\end{equation} 
where $a$ and $k$ denote amplitude and wave number, respectively. The dispersion relation is therefore
\begin{equation} \label{eq:omega-k}
\omega^2 = c_{s,0}^2 k^2.
\end{equation}
Based on Eqs.~\eqref{eq:eom}, \eqref{eq:rhoP} and the dispersion relation \eqref{eq:omega-k}, the expression of fluid velocity can be written as
\begin{equation} \label{eq:velo}
V = \frac{c_{s,0}}{\rho_0} a \cos(ky - \omega t). 
\end{equation}
Comparing Eq.~\eqref{eq:rhoP} and Eq.~\eqref{eq:velo} gives the relation between pressure fluctuation and fluid velocity:
\begin{equation} \label{eq:Pp-V}
P^{\prime} = \rho_0 c_{s,0} V.
\end{equation}
We recall that, for adiabatic oscillations, pressure and temperature fluctuations are related via
\begin{equation} \label{eq:TpandPp}
\frac{T^{\prime}}{T_0} = \nabla_{\rm ad,0} \frac{P^{\prime}}{P_0},
\end{equation}
with $\nabla_{\rm ad} = (\partial \ln T / \partial \ln P)_{\rm ad}$ being the adiabatic temperature gradient. The relation between the temperature fluctuation and the fluid velocity is given by
\begin{equation} \label{eq:TpandV}
T^{\prime} = \frac{\rho_0 c_{s,0} T_0 \nabla_{\rm ad,0}}{P_0} V.
\end{equation}
For additional information about relevant discussion and derivations, see \citet{1987flme.book.....L} chapter 64 and \citet{2010aste.book.....A} chapter 3.1.4.

   We have now obtained the relationship between $\delta L/L_0$ and $\delta T$ (the temperature fluctuation at constant optical depth), as well as the relationship between $V$ and $T^{\prime}$ (the temperature fluctuation at constant geometric depth). The next step is to link these two temperature fluctuations. Considering only first order perturbations, the temperature at given optical depth $\tau$ at any given time can be separated as
\begin{equation}
T(\tau,t) = T_0(\tau) + \delta T(\tau,t).
\end{equation}
At fixed geometric depth near the photosphere, the optical depth varies with time because of the time-dependent nature of convection. Therefore, the temperature at fixed geometric depth, if expressed as a function of $\tau$, reads
\begin{equation}
\begin{aligned}
T(\tau + d\tau,t) = T(\tau,t) + \frac{\partial T}{\partial \tau} d\tau
= T_0(\tau) + \delta T(\tau,t) + \frac{\partial T}{\partial \tau} d\tau.
\end{aligned}
\end{equation}
Because $T_0(\tau)$ represents the equilibrium state, the Eulerian perturbation to temperature is therefore
\begin{equation} \label{eq:TpanddT}
T^{\prime}(\tau,t) = T(\tau + d\tau,t) - T_0(\tau) 
= \delta T(\tau,t) + \frac{\partial T}{\partial \tau} d\tau.
\end{equation}
This equation demonstrates the relationship between two different kinds of perturbation, it can be expanded further by analysing the term $d\tau$. In the equilibrium state, the optical depth is, by definition,
\begin{equation} \label{eq:tau}
\tau = \int_{-\infty}^{y_0} \alpha_0 \; dy,
\end{equation}
where $y$ is the geometric depth as before and $\alpha_0$ is the mean absorption coefficient. 
Recalling that at a given time $t$, $\tau + d\tau$ corresponds to the same geometric depth $y_0$, we have
\begin{equation} \label{eq:tau+dtau}
\tau + d\tau = \int_{-\infty}^{y_0} \alpha(t) \; dy.
\end{equation}
Subtracting Eq.~\eqref{eq:tau} from Eq.~\eqref{eq:tau+dtau} gives
\begin{equation} \label{eq:dtau}
d\tau = \int_{-\infty}^{y_0} \delta\alpha(t) \; dy,
\end{equation}
which relates the perturbation of the absorption coefficient at constant $\tau$ to the change in optical depth at fixed geometric depth. The value of the absorption coefficient, however, depends on the opacity source of the plasma in a complex way. As such, there is no simple analytical function to describe the relationship between $\alpha$ and the thermodynamical quantities. Nevertheless, given the fact that the $\rm H^-$ opacity is the dominant source of opacity near the solar photosphere, we adopt the simplification that the mass absorption coefficient consists only of $\rm H^-$ opacity: $\kappa_{\rm H^-}$ (units cm$^2$g$^{-1}$). Here we adopt a power-law fit of $\kappa_{\rm H^-}$ (\citealt{2004sipp.book.....H} Eq.~4.65), which gives reasonable results in our range of interest ($3000 \lesssim T \lesssim 6000$ K; $10^{-10} \lesssim \rho \lesssim 10^{-5} \; \rm g/cm^3$; hydrogen mass fraction of around 0.7; metal mass fraction $0.001 \lesssim Z \lesssim 0.03$):
\begin{equation} \label{eq:kappaH-}
\begin{aligned}
\kappa_{\rm H^-} &\simeq 2.5 \times 10^{-31} (Z/0.02) \rho^{1/2} T^9 \qquad 
\rm cm^2/g,
\\
\alpha \simeq \rho\kappa_{\rm H^-} 
&\simeq 2.5 \times 10^{-31} (Z/0.02) \rho^{3/2} T^9 \qquad \rm cm^{-1}.
\end{aligned}
\end{equation}

The perturbation of the absorption coefficient is then
\begin{equation} \label{eq:dalpha}
\delta\alpha \simeq \alpha_0 
\left( \frac{3 \delta\rho}{2\rho_0} + \frac{9 \delta T}{T_0} \right),
\end{equation}
where $\delta\rho$ is the perturbation of density at fixed optical depth. As indicated by 3D surface convection simulations, the magnitude of $\delta\rho$ and $\rho^{\prime}$ are similar around photosphere (Fig.~3 of \citealt{2013A&A...560A...8M}). Hence, using also Eqs.~\eqref{eq:energyeq_ad}, \eqref{eq:Pp-V} and \eqref{eq:TpandV}, we have
\begin{equation} \label{eq:drhoanddT}
\delta\rho \simeq \rho^{\prime} = 
\frac{P_0}{c_{s,0}^2 \nabla_{\rm ad,0} T_0} T^{\prime}.
\end{equation}
Substituting Eqs.~\eqref{eq:dalpha} and \eqref{eq:drhoanddT} into Eq.~\eqref{eq:dtau} yields
\begin{equation} \label{eq:dtauexpan}
d\tau \simeq \int_{-\infty}^{y_0} 
\alpha_0 \left(\frac{3P_0}{2\rho_0 c_{s,0}^2 \nabla_{\rm ad,0} T_0} T^{\prime} +  
\frac{9}{T_0} \delta T \right) \; dy.
\end{equation} 
As the absorption coefficient $\alpha_0$ increases rapidly when moving from the upper atmosphere to photosphere (Eq.~\eqref{eq:kappaH-}), the main contribution to the right hand side of Eq.~\eqref{eq:dtauexpan} comes from a thin layer just above $y_0$. Therefore, Eq.~\eqref{eq:dtauexpan} can be approximated by
\begin{equation} \label{eq:dtaufinal}
d\tau \simeq \tau \left(\frac{3P_0}{2\rho_0 c_{s,0}^2 \nabla_{\rm ad,0} T_0} 
T^{\prime} +  \frac{9}{T_0} \delta T \right)_{\tau}.
\end{equation}
Under the assumption of the Eddington grey atmosphere and LTE, the temperature stratification is
\begin{equation}
T(\tau) = T_{\rm eff} \left( \frac{3}{4}\tau + \frac{1}{2} \right)^\frac{1}{4},
\end{equation}
which is the so-called Eddington $T-\tau$ relation. Making use of the Eddington $T - \tau$ relation and plugging Eq.~\eqref{eq:dtaufinal} into Eq.~\eqref{eq:TpanddT}, we have
\begin{equation}
T^{\prime} \simeq \delta T + \frac{3}{16} T_{\rm eff,0} 
\left( \frac{3}{4}\tau + \frac{1}{2} \right)^{-\frac{3}{4}}
\tau \left(\frac{3P_0}{2\rho_0 c_{s,0}^2 \nabla_{\rm ad,0} T_0} T^{\prime} +  
\frac{9}{T_0} \delta T \right)_{\tau}.
\end{equation}
Evaluating the equation above at $\tau = 2/3$ gives the relationship between $T^{\prime}$ and $\delta T$ at the optical surface:
\begin{equation} \label{eq:TpanddTsurf}
\left( 1 - \frac{3P_0}{16 \rho_0 c_{s,0}^2 \nabla_{\rm ad,0}} \right)_{\tau=\frac{2}{3}} T^{\prime}(\tau=2/3,t) 
\simeq \frac{17}{8} \delta T(\tau=2/3,t).
\end{equation}
The ratio between $T^{\prime}$ and $\delta T$ at $\tau = 2/3$ computed based on Eq.~\eqref{eq:TpanddTsurf} is approximately 3.2 for our solar atmosphere model, in reasonable agreement with the corresponding result evaluated directly from simulation data, which is approximately 2.3 (see also \citealt{2013A&A...560A...8M} Sect.~4.1).

  Finally, combining Eqs.~\eqref{eq:dL-T}, \eqref{eq:TpandV} and \eqref{eq:TpanddTsurf} gives the relationship between the luminosity and velocity amplitudes:
\begin{equation} \label{eq:ppmandv}
\frac{\delta L}{L_0} \simeq \frac{32}{17} \times 10^6 
\left( \frac{\rho_0 c_{s,0} \nabla_{\rm ad,0}}{P_0} - \frac{3}{16c_{s,0}} \right)_{\tau = \frac{2}{3}} V(\tau=2/3).
\end{equation}
Eq.~\eqref{eq:ppmandv} demonstrates that, to first order, the ratio between relative luminosity variation and photosphere velocity depends only on the equilibrium state of the thermodynamic quantities. We are now equipped to address the question put forward at the end of Sect.~\ref{sec:3Dmodel}: it is thus demonstrated that 3D surface convection simulations do have the potential to reliably predict the luminosity and velocity amplitude ratio, because the ratio does not depend on the luminosity or velocity amplitude nor on any other term that is subject to overestimation by our box-in-a-star models.

  We note that a number of approximations and simplifications have been employed when deriving Eq.~\eqref{eq:ppmandv}. Itemized, these assumptions are:
\begin{itemize}
\item  The Eddington-Barbier approximation

\item  Local thermodynamic equilibrium

\item  Grey atmosphere

\item  Convective velocities are regarded as the ``equilibrium state'', such that fluid velocity $V$ consists only of an oscillation component and is small compared to the sound speed

\item  Spatially homogeneous medium

\item  The Cowling approximation

\item  Adiabatic oscillations

\item  $\rm H^-$ is the only source of opacity in the stellar photosphere, such that it can be represented by a power law $\kappa_{\rm H^-} \propto \rho^{1/2} T^9$

\item  The magnitude of $\delta\rho$ and $\rho^{\prime}$ are similar around photosphere
\end{itemize}
Some of these assumption, such as the grey atmosphere and spatially homogeneous medium assumptions, are obviously not correct in the near-surface regions. Therefore, the analysis above is only to illustrate that 3D simulations are capable of providing a reliable luminosity and velocity amplitude ratio. We emphasise, however, that Eq.~\eqref{eq:ppmandv} is not used to calculate the ratio between luminosity and velocity amplitude; rather, we evaluate luminosity variation and radial velocity directly from 3D simulations that do not rely on these assumptions.

\section{Evaluating luminosity amplitude} \label{sec:bol}

\subsection{Intrinsic bolometric amplitude} \label{sec:Abol}

\begin{figure*}
\subfigure[]{
\begin{overpic}[width=0.49\textwidth]{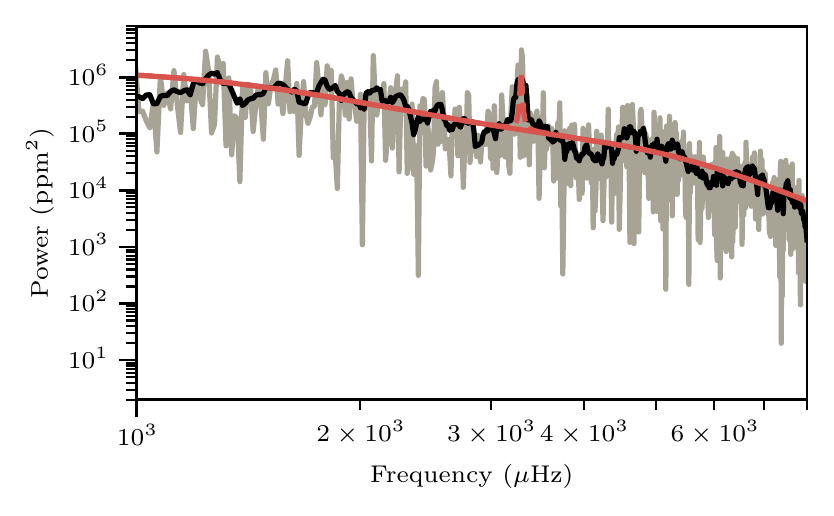}
\end{overpic}
\label{fig:sunppm}
}
\subfigure[]{
\begin{overpic}[width=0.49\textwidth]{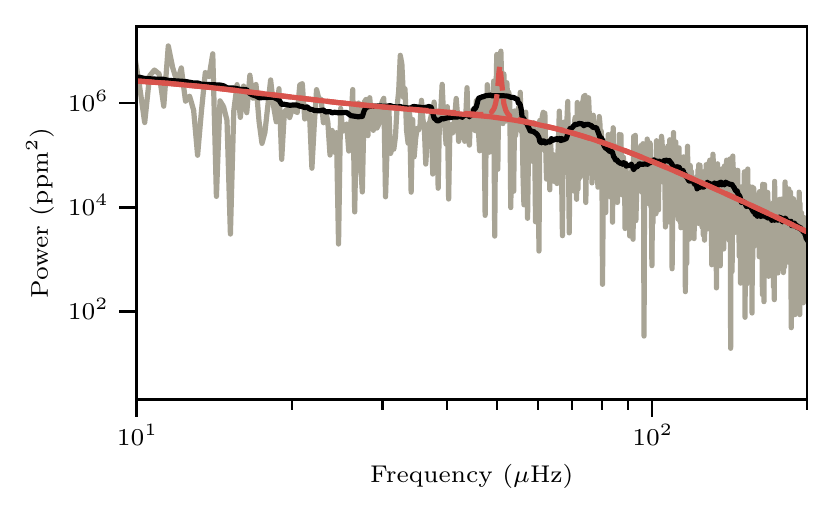}
\end{overpic}
\label{fig:t50g27m00ppm}
}
\caption{\ref{fig:sunppm}: Grey line is the luminosity power spectrum computed from the solar simulation, and the black curve is the result after smoothing with a running mean with width equals to 100 $\rm \mu Hz$. Red solid line represents the granulation background of the solar simulation, as modelled based on Eq.~\eqref{eq:PSgran}. A Lorentzian fit to the dominant simulation mode is also shown in red dashed line. 
\ref{fig:t50g27m00ppm}: Luminosity power spectrum and granulation background predicted from our $\epsilon$ Tau simulation. The black curve is obtained by smoothing the luminosity power spectrum with a running mean with width equals to 10 $\rm \mu Hz$.}
\label{fig:ppm}
\end{figure*}

  Intrinsic bolometric flux is an output quantity from the 3D stellar surface convection simulations. It is computed from the radiative transfer calculations performed at each time step of the simulation (Sect.~\ref{sec:3Dmodel}). The theoretical bolometric flux as a function of time, which is analogous to the intrinsic light curve of star, is converted to $\delta L / L_0$ according to Eq.~\eqref{eq:dL_def}. This is then transformed to the (oscillation) frequency domain using a Lomb-Scargle Periodogram algorithm \citep{1976Ap&SS..39..447L,1982ApJ...263..835S} to obtain the luminosity power spectrum shown in Fig.~\ref{fig:ppm} (grey lines). A general trend of the luminosity power spectra is that the luminosity power is higher at low frequencies and decreases with increasing frequency. In the solar case, a peak located around 3.3 mHz is clearly seen in the spectrum. This feature is associated with surface convection (granulation), where up- and downflows shift the location of the optical surface, producing fluctuations in bolometric flux. The peak around 3.3 mHz is caused by the main oscillation mode of the simulation box. Acoustic waves naturally excited in the simulation domain periodically change the location of optical surface, leading to coherent variations in bolometric flux. The variation due to granulation happens on all time-scales and thus provides the background signal in the power spectrum; for an even longer time-sequence this granulation signal becomes more and more smooth, making it easier to discern the frequencies of the oscillation modes. 
  
  In order to obtain the luminosity amplitude for the simulation mode, it is necessary to filter out the contribution from granulation. At a given spatial position near the photosphere, granulation emerges, evolves, and disappears with a typical time scale $t_{\rm gran}$. Having the insight that granulation (essentially surface velocity field) is constantly evolving, \citet{1985ESASP.235..199H} proposed that the autocorrelation function of stellar granulation can be described by exponential function $\exp(-t/t_{\rm gran})$. That is, the correlation between granulation at moments $t_0$ and $t_0 + t$ decreases exponentially with increasing time interval. Because the autocorrelation of a signal corresponds to the Fourier transform of its power spectrum, the power spectrum of granulation background is the Fourier transform of
\begin{equation}
\mathcal{V}(t) = 
\begin{cases} 
\mathcal{V}_0 e^{-t/t_{\rm gran}} & (t \geq 0)
\\ 
0 & (t<0)
\end{cases},
\end{equation}
which is
\begin{equation} 
\mathcal{B}(\nu) = \frac{C \mathcal{V}_0^2 t_{\rm gran}}{1 + (2\pi\nu t_{\rm gran})^2}.
\end{equation}
This is a Lorentzian function. The term $\mathcal{V}_0$ is the velocity amplitude associated with granulation, $\nu$ is cyclic frequency and $C$ is a normalization constant such that the power spectrum satisfies the Parseval theorem (see also Eq.~5 of \citealt{2017ApJ...835..172L}). In recognition of this, we modelled the oscillation background caused by granulation with the sum of one or more (generalized) Lorentzian profiles:\footnote{Strictly speaking, Eq.~\eqref{eq:PSgran} is the sum of generalized Lorentzian profiles because $a_{3,i}$ is a free parameter rather than fixed to 2. Here we refer them as Lorentzian profiles for simplicity.}
\begin{equation} \label{eq:PSgran}
\mathcal{B}(\nu) = \sum_{i=1}^{N} \frac{2\sqrt{2}}{\pi} 
\frac{a_{1,i}^2 / a_{2,i}}{ 1 + (\nu / a_{2,i})^{a_{3,i}} },
\end{equation}
which is similar to the functional forms commonly applied to real observational data (e.g.~\citealt{2017ApJ...835..172L,2020MNRAS.495.2363L}). Here, the free parameters are $\{ a_{1,i}, a_{2,i}, a_{3,i} \}$ and $N$ is the number of Lorentzian components. As multiple Lorentzian components are often employed to achieve a better fit to the observed stellar background, the interpretation is that in real stars, there exists more than one granulation scale as well as contributions from stellar activity.
Nevertheless, the true analytical form of stellar granulation background remains elusive \citep{2020MNRAS.tmp.3425L}. Therefore, we test granulation background models with one, two and three components ($N=1,2,3$) by computing the Bayesian evidence (marginal likelihood) for each model. The Bayesian evidence $p(D\vert\mathcal{M})$, which is the probability of the power spectrum data $D$ given a granulation model $\mathcal{M}$, is frequently used to evaluate the relative probability of the models for given data. We find that for both the Sun and $\epsilon$ Tau, the Bayesian evidence of the single-component background model is significantly smaller than multi-component models while $p(D\vert\mathcal{M})$ of $N=2$ and $N=3$ models are not significantly different. 
Our Bayesian approach therefore shows that given the theoretical luminosity spectrum, the granulation background is better described by multi-component Lorentzian profiles. In this study, we choose the two-component ($N=2$) model, because it performs equally well as the three-component one but involves fewer free parameters. We refer the readers to \citet{2020MNRAS.tmp.3425L} for a detailed examination of different granulation background models against results from 3D surface convection simulations.

  The granulation background fitting is performed with the parallel tempering MCMC (Markov chain Monte Carlo) algorithm of \citet{2016MNRAS.455.1919V}. The best-fitting results are demonstrated in red solid lines in Fig.~\ref{fig:ppm}. The amplitude of the granulation background at the frequency of the dominant simulation mode is $343.6 \pm 14.7$ ppm for the solar simulation and $726.0 \pm 31.2$ ppm for the $\epsilon$ Tau case. Uncertainties presented here are returned from the MCMC samples, representing the statistical errors associated with the background fitting. 
We then subtracted the power spectrum with the best fitting background. The bolometric oscillation amplitude is determined by taking the square root of the peak value (corresponds to the value at the frequency of the dominant simulation mode) of the subtracted spectrum, which is $1714.0 \pm 3.0$ ppm for the solar simulation and $3070.3 \pm 7.4$ ppm for $\epsilon$ Tau (also tabulated in Table \ref{tb:result}). We note that another frequently used method to extract the oscillation amplitude is to fit a parametric model to the peak region, then take the peak value of the fitted curve. However, in our solar simulation, the width of the simulation mode, which is connected to the mode damping rate, is on the order of $10-100$ $\mu$Hz (see also Table 1 of \citealt{2019A&A...625A..20B}), being much larger than the width of solar $p$-modes. The reason is that the simulations modes have much less mode mass than stellar $p$-modes. Therefore, to avoid further complications about the reliability of the width, we measure only the peak amplitude at the frequency of the dominant simulation mode in our analysis.

\subsection{The conversion factor between intrinsic and measured luminosity amplitude} \label{sec:cP-bol}

\begin{figure*}
\subfigure{
\begin{overpic}[width=0.99\textwidth]{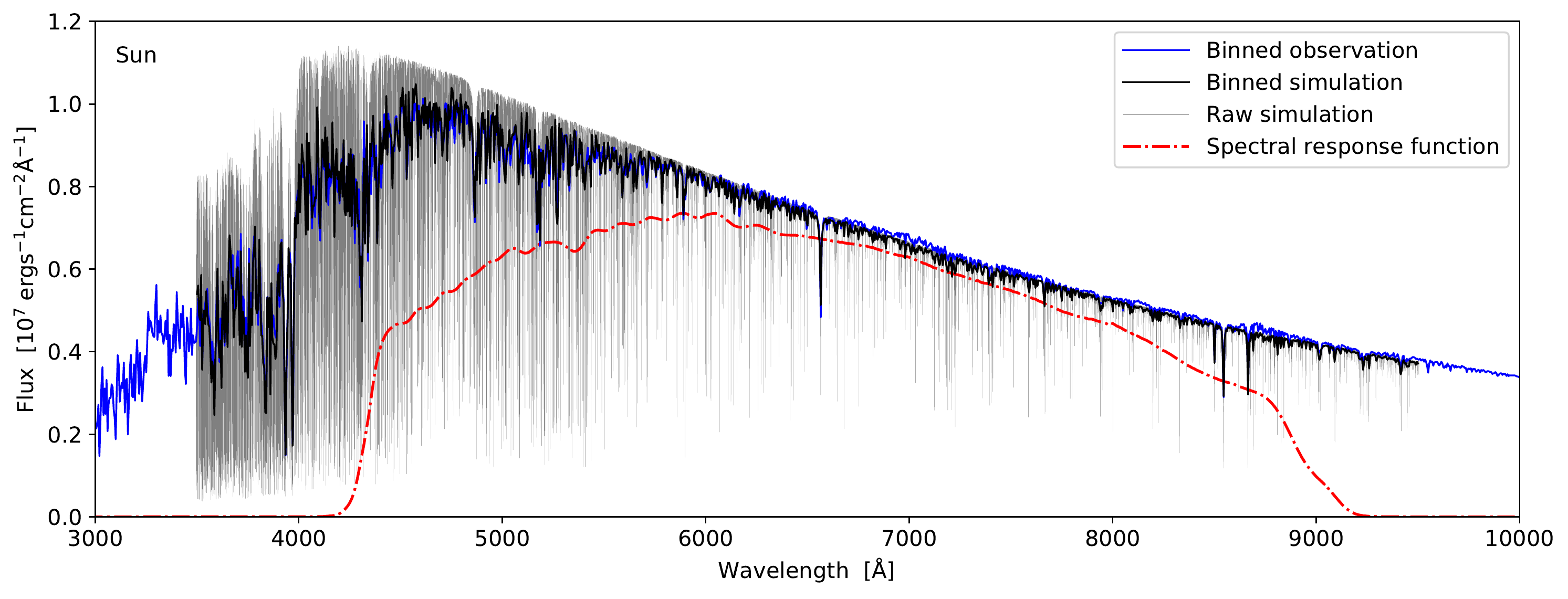}
\end{overpic}
}
\subfigure{
\begin{overpic}[width=0.99\textwidth]{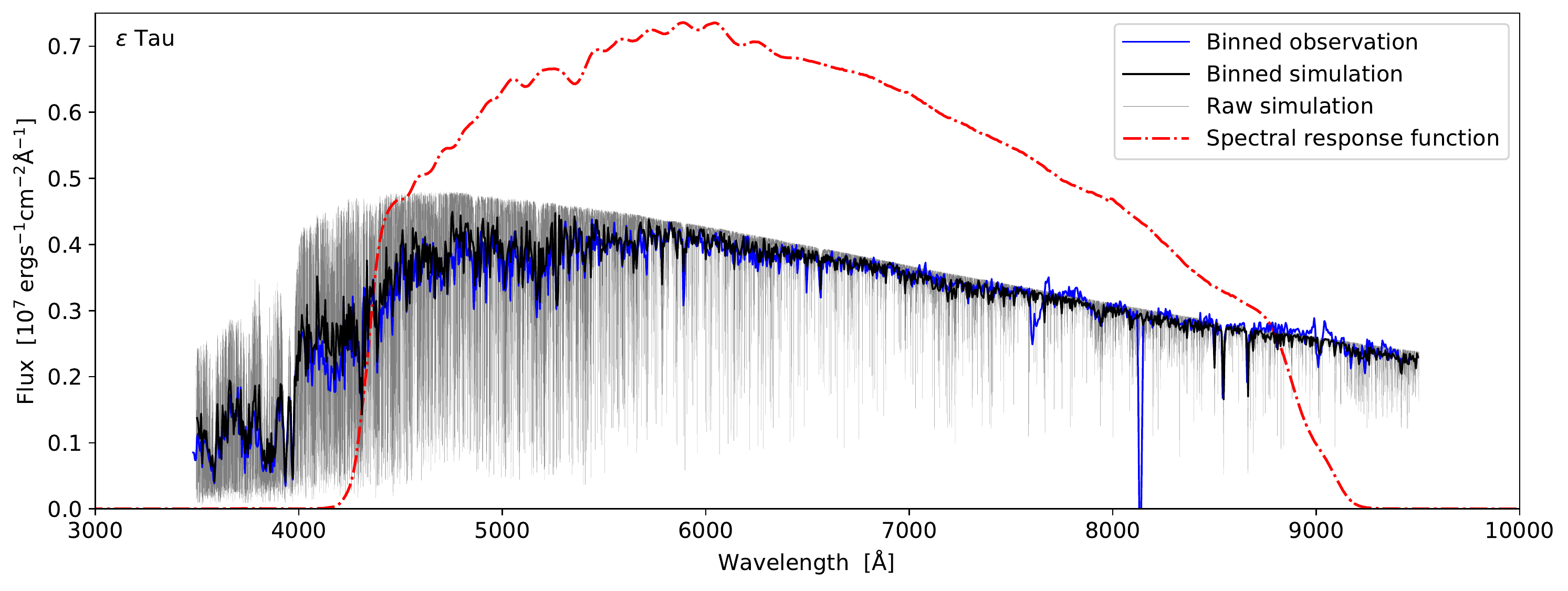}
\end{overpic}
}
\caption{\textit{Upper panel}: Thin grey line is the emergent flux between 3500 and 9500 {\AA} computed based on one example snapshot of our 3D solar model atmosphere (at a resolution of 15\,k\ms). The re-sampled synthetic spectrum using a 5 {\AA} wavelength bin is depicted with the black line. The observed high resolution solar flux spectrum of \citet{2005MSAIS...8..189K} is also binned every 5 {\AA} (blue line) in order to facilitate comparison between simulation and observation. Red dash-dotted line is the mean spectral response function of \textit{Kepler} averaged from its 84 channels \citep{2016ksci.rept....9T}.
\textit{Lower panel}: Similar to the upper panel, but the result for $\epsilon$ Tau. Because the observed $\epsilon$ Tau spectrum of \citet{2004ApJS..152..251V} lacks an absolute flux level, we normalise the original observation data such that its magnitude generally matches the 4976\,K (the reference effective temperature of $\epsilon$ Tau) black body spectrum between 7000 and 9000\,{\AA}. We note that data shown in this figure is not used to calculate $c_{P-\rm bol}$. Instead, the synthetic spectra that enter into Eq.~\eqref{eq:cp-bol} have a lower wavelength resolution (1\,nm, see text).
}
\label{fig:spectrum}
\end{figure*}

\begin{table}
	\centering
	\caption{Conversion factor $c_{P-\rm bol}$ computed for \textit{Kepler} for the Sun and $\epsilon$ Tau. ``3D'' , ``ATLAS9'' and ``Planck'' represent the choice of model atmosphere (3D model atmosphere in this work, ATLAS9 model atmosphere, and black body respectively) applied in the computation. The ``ATLAS9'' and ``Planck'' results are obtained by interpolating the data provided in \citet{2019MNRAS.489.1072L}.}
	\label{tb:cp-bol}
	\begin{tabular}{cccc} 
    \toprule[2pt]
Star           & 3D   & ATLAS9 & Planck
\\
	\midrule[1pt]
Sun            & 0.92 & 0.94   & 0.98
\\
$\epsilon$ Tau & 0.68 & 0.79   & 0.87
\\
	\bottomrule[2pt]
	\end{tabular}
\end{table}

  Space-based missions such as CoRoT, \textit{Kepler} and TESS measure starlight in a certain band-pass. What is measured from these space-based observations is stellar flux in certain wavelength ranges, rather than the intrinsic bolometric stellar flux. In order to connect the luminosity amplitude provided by 3D simulations with observables, it is necessary to quantify the conversion factor between intrinsic and measured luminosity amplitude (also called bolometric correction factor for luminosity amplitude) which is defined as
\begin{equation} \label{eq:Abol-AP}
A_{\rm bol} = c_{P-\rm bol} A_P,
\end{equation}
where $c_{P-\rm bol}$ is the conversion factor. The term $A_{\rm bol}$ and $A_P$ are intrinsic luminosity amplitude and luminosity amplitude measured in a certain band-pass respectively. The conversion factor was first investigated by \citet{2009A&A...495..979M} for CoRoT and \citet{2011A&A...531A.124B} for \textit{Kepler}. Recently, \citet{2019MNRAS.489.1072L} quantified $c_{P-\rm bol}$ values for CoRoT, \textit{Kepler} and TESS across the HR diagram based on a grid of ATLAS9 model fluxes \citep{2003IAUS..210P.A20C}. Here we follow their theoretical formulation, but we calculate the flux spectrum from our 3D models to obtain self-consistent conversion factors for the two stars investigated in this work. For radial oscillations, the expression of $c_{P-\rm bol}$, as given in \citet{2019MNRAS.489.1072L}, is 
\begin{equation} \label{eq:cp-bol}
c_{P-\rm bol} = \frac{4\int \; \mathcal{T}_P(\lambda) F(\lambda) d\lambda}
{T_{\rm eff} \int \; \mathcal{T}_P(\lambda) 
\frac{\partial F(\lambda)}{\partial T_{\rm eff}} d\lambda},
\end{equation}
where $\lambda$ is wavelength. Note that the stellar spectral flux $F(\lambda)$ also depends on basic stellar parameters i.e.~$T_{\rm eff}$, $\log g$ and [Fe/H]. The instrumental transfer function $\mathcal{T}_P(\lambda)$ is connected with the spectral response function $\mathcal{S}_{\lambda}$ via $\mathcal{T}_P(\lambda) = \mathcal{S}_{\lambda} / (hc/\lambda)$, where the latter represents the band-pass of the instrument. In this study we choose the \textit{Kepler} spectral response function as an example.

  The stellar spectral flux $F(\lambda)$ in Eq.~\eqref{eq:cp-bol} is computed using the 3D radiative transfer code \textsc{scate} \citep{2011A&A...529A.158H}.
\textsc{scate} solves the 3D, time-dependent radiative transfer problem for both spectral lines and the background continuum. The computation is carried out under the LTE assumption, but with the ability to include isotropic continuum scattering in opacities and source functions. Simulation snapshots generated from the \textsc{Stagger} code are input models in \textsc{scate}. The equation-of-state, continuum absorption and scattering coefficients, and the pre-tabulated line opacities adopted in \textsc{scate} are identical to those used in \textsc{Stagger}, thereby ensuring full consistency between the 3D surface convection simulations and 3D LTE line formation calculations. For more information about the code and the numerical method therein, we refer the readers to \citet{2010A&A...517A..49H,2011A&A...529A.158H}. 

  Here we use the spectrum synthesis mode of \textsc{scate}. In this scenario, the code delivers the angle-resolved surface fluxes for many wavelength points while treating scattering as pure absorption; test calculations reveal that for our target stars and for the wavelengths of interest here, this is an excellent approximation. At a given wavelength, continuum opacities are computed on-the-fly based on the microphysics and 3D atmosphere models mentioned above; line opacities are read from the pre-tabulated opacity sampling data. Specific intensities are calculated by tilting the simulation domain to represent five different polar angles $\theta$, and rotated to yield four equidistant azimuthal angles $\phi$, for a total of 20 rays. The emergent flux is finally computed through integration of the specific intensities on a Gauss-Legendre quadrature in the polar direction, and trapezoidal integration in the azimuthal direction. 
  We compute the emergent flux between 350 and 950\,nm, covering the entire spectral response function of \textit{Kepler}, in steps of 1\,nm. In order to reduce the computational cost, we compute the spectral energy distribution only for a subset of the simulation sequence covering four periods of the dominant simulation mode. This corresponds to 40 snapshots in the solar simulation and 55 snapshots in the red giant simulation. 
  The spectral flux distributions computed based on one example snapshot of our 3D solar and $\epsilon$ Tau model atmosphere are depicted in Fig.~\ref{fig:spectrum}, together with the spectral response function $\mathcal{S}_{\lambda}$ for \textit{Kepler} which is taken from \citet{2016ksci.rept....9T}. From Fig.~\ref{fig:spectrum} we can see that for both stars, the predicted spectral flux distributions agree reasonably well with observations, both in magnitude and in overall trend.
  
  The time-averaged (average over all selected snapshots) spectral flux $F(\lambda)$ is used to calculate $c_{P-\rm bol}$ through Eq.~\eqref{eq:cp-bol}. The derivative term $\partial F(\lambda) / \partial T_{\rm eff}$ can be evaluated from the same simulation snapshots. More specifically, by synthesising the flux spectrum for each individual simulation snapshot, we can obtain $F(\lambda)$ as a function of $T_{\rm eff}$ because different simulation snapshots correspond to different bolometric flux, hence effective temperature. This then permits the numerical evaluation of $\partial F(\lambda) / \partial T_{\rm eff}$ at the reference effective temperature of the star. 
  The $c_{P-\rm bol}$ values computed based on the aforementioned simulation configuration are presented in Table \ref{tb:cp-bol}. In the case of $\epsilon$ Tau, the conversion factor calculated from the 3D model is 0.68, more than 10\% less than the corresponding ATLAS9 result given in \citet{2019MNRAS.489.1072L}. 
  We have verified for both the Sun and $\epsilon$ Tau that our $c_{P-\rm bol}$ value is robust, as (1) increasing the number of polar angles in the radiative transfer calculation from five to ten has negligible effect on the final $c_{P-\rm bol}$ value; and (2) neither increasing the wavelength resolution from 1 nm to 1 {\AA} nor doubling the simulation time sequence in the flux spectrum calculation changes the outcome.
  Different model atmospheres is a likely cause of the discrepancy in $c_{P-\rm bol}$, as $c_{P-\rm bol}$ computed from the black body spectrum also differs clearly from the one computed from ATLAS9, as seen in Table \ref{tb:cp-bol}. 

  It is worth noting that in principle the conversion factor $c_{P-\rm bol}$ is not strictly a constant because the flux emitted by a star fluctuates with time. The fluctuation is caused by stellar granulation and oscillation for solar-type stars with the contribution from granulation generally being much larger (\citealt{2014A&A...570A..41K}). According to \citet{2014A&A...570A..41K}, the measured solar bolometric granulation amplitude is $A_{\rm gran} = 41$ ppm, corresponding to a fluctuation of approximately 0.06 K in effective temperature. Assuming a black body spectrum, the 0.06 K fluctuation in $T_{\rm eff}$ will result in a relative change of less than $10^{-5}$ in $c_{P-\rm bol}$. For $\epsilon$ Tau, the bolometric granulation amplitude, estimated from the scaling relation $A_{\rm gran} \propto (g^2 M)^{-1/4}$ (\citealt{2014A&A...570A..41K} Eq.~5), is roughly 250 ppm, corresponding to a $\sim 0.3$ K temperature fluctuation and a relative change in $c_{P-\rm bol}$ of less than $10^{-4}$. Therefore, regarding the conversion factor as a constant is a suitable approximation.

\section{Evaluating radial velocity amplitude} \label{sec:RV}

  In the spectroscopic method of measuring stellar oscillations, variations in the radial (i.e.~line-of-sight) velocity near the stellar photosphere are quantified by analysing the Doppler shift of certain spectral lines. Two representative efforts based on this method are the Birmingham Solar Oscillations Network (BiSON, \citealt{1996SoPh..168....1C}) and Stellar Oscillations Network Group (SONG, \citealt{2006MmSAI..77..458G}). As implied by its name, BiSON focus solely on helioseismology: it detects solar oscillations using the Doppler shift of the disk-integrated solar potassium (\ion{K}{i}) 7698 {\AA} line. The spectrograph operates by imposing a magnetic field on a sample of potassium gas; the anomalous Zeeman effect produces line splitting where the $\sigma_-$ and $\sigma_+$ line components located in the blue and red wings of the solar \ion{K}i line exhibit circular polarisation with opposite orientation. 
  Incident sunlight fed through the instrument passes through a linear polarizer and a quarter-wave plate, which induces circular polarization that can be rapidly switched between left- and right-handedness in order to produce resonant scattering with either line component.
  This allows the measurement of the relative intensity between the blue and red wing, which reflects the Doppler shift of solar \ion{K}{i} line that is caused by the velocity field of the solar surface. A thorough explanation of the BiSON instrumentation and observation technique can be found in \citet{1996SoPh..168....1C}.
  
  In contrast, SONG measures the radial velocity signal simultaneously from a large number of spectral lines through a traditional echelle spectrograph that covers the wide spectral range 4400--6900\,\AA\ \citep{2017ApJ...836..142G}. The incident starlight passes through an iodine cell, which superimposes on the stellar spectrum a large number of weak absorption lines; these act as a highly accurate simultaneously recorded wavelength reference. The overall Doppler shift is inferred by cross-correlating the observed spectrum with a reference spectrum that was recorded without the iodine cell. Because radial velocity quantified in this manner takes many spectral lines into account, the result reflects the mean velocity of the photosphere rather than the velocities at the specific heights where certain lines are formed. A more detailed description of the iodine cell method is given by \citet{1996PASP..108..500B}, and its application to SONG is described in detail by \citet{2013MNRAS.435.1563A}.
  To make theoretical predictions as consistent as possible with observations, we extract the radial velocity from our 3D atmosphere models through line formation calculations with a method resembling the BiSON and SONG observational setups in the following subsections.

\subsection{Simulating the BiSON radial velocity signal} \label{sec:BiSON}

\begin{figure}
\begin{overpic}[width=\columnwidth]{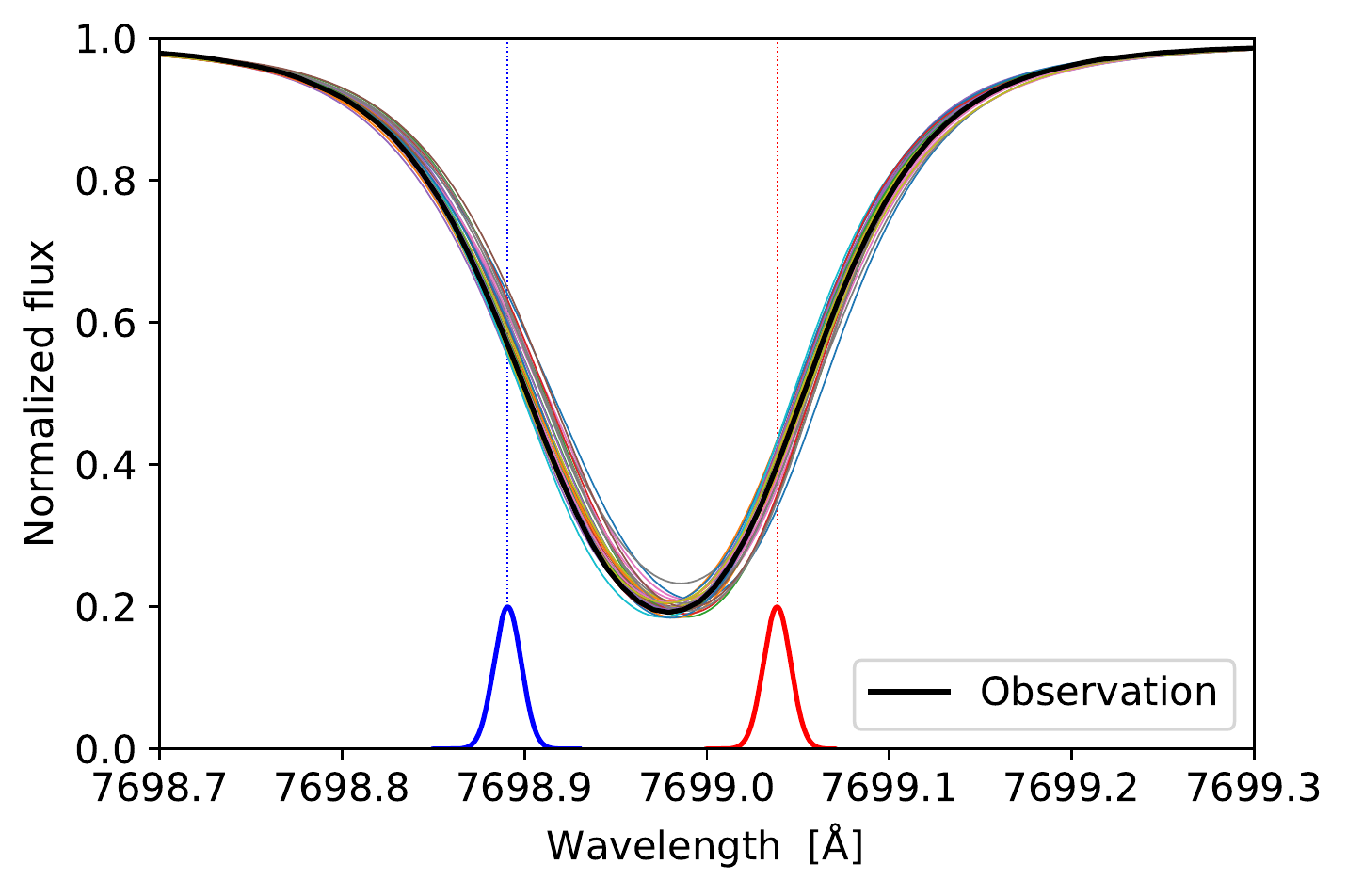}
\end{overpic}
\caption{Spatially averaged \ion{K}{i} line profiles predicted from our 3D non-LTE line formation calculations. The line formation calculation is done for every simulation snapshot, but only one in every 100 snapshots are shown here to avoid over-crowded figure. The observed solar \ion{K}{i} line profile is plotted in black line for comparison \citep{1999SoPh..184..421N}. The effects of gravitational redshift (633\,\ms\ for the Sun, \citealt{2008A&A...492..199D}) are included in both the theoretical and the observed line profiles. The blue and red Gaussian profiles schematically represent the two laboratory potassium lines used in BiSON, whose central wavelengths are marked by vertical dotted lines.}
\label{fig:KIline}
\end{figure}

\begin{figure}
\begin{overpic}[width=\columnwidth]{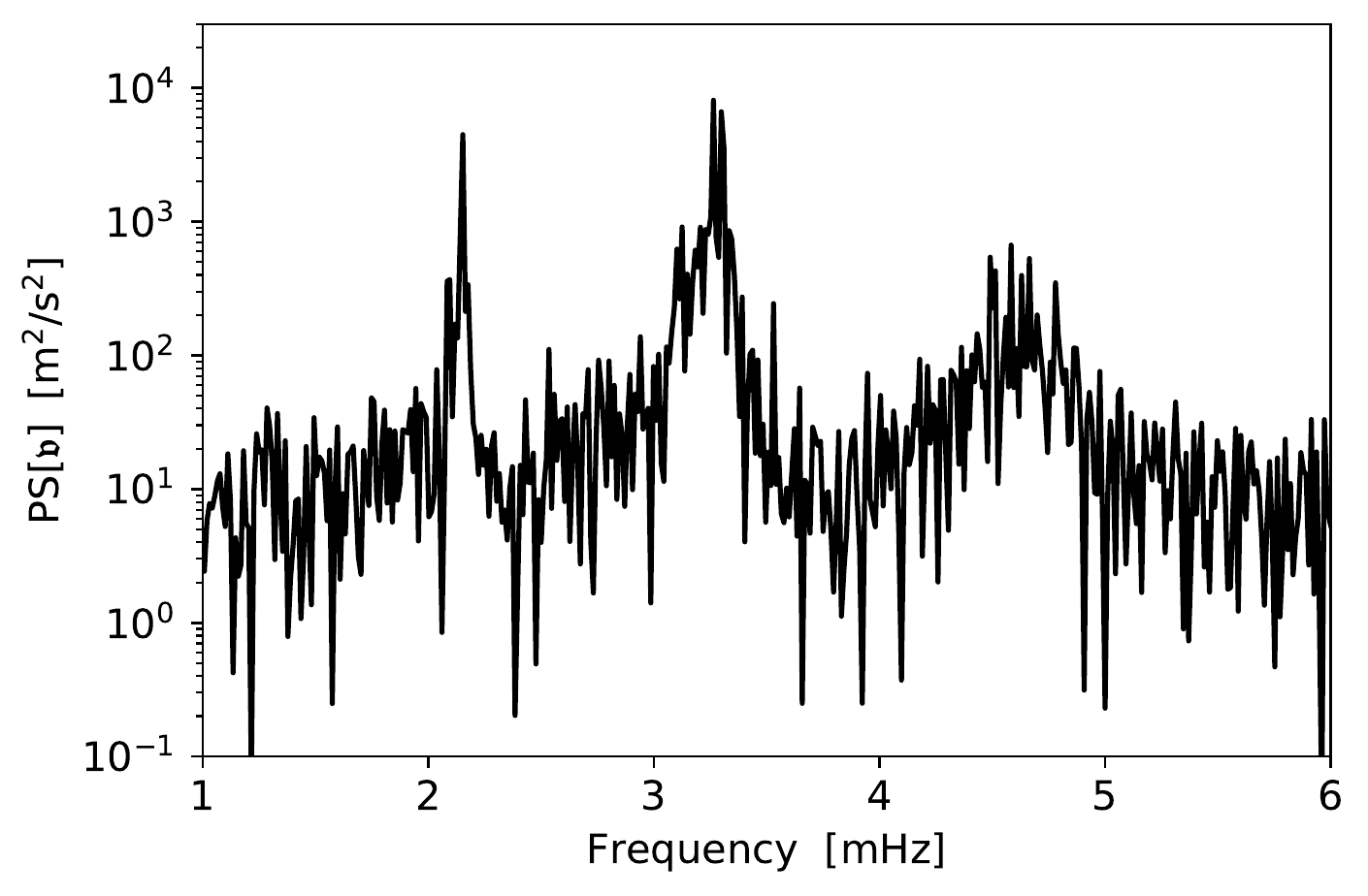}
\end{overpic}
\caption{The power spectrum of radial velocity amplitude, with three peaks correspond to three simulation modes. The power spectrum here should not be confused with the grey line in Fig.~\ref{fig:sunPS}, the plot here reflects radial velocity variation in the \ion{K}{i} line forming regions while the latter is the fluctuation of vertical component of fluid velocity near the photosphere.
}
\label{fig:PSRV}
\end{figure}

  Although 3D hydrodynamic simulations are able to realistically predict the convective velocities throughout the simulation domain \citep[e.g.][]{2000A&A...359..729A,2009LRSP....6....2N,2013A&A...554A.118P}, what BiSON measures is a radial velocity as imprinted on a particular spectral line, which is connected but not equivalent to the fluid velocity given by 3D simulations. In practice, spectral lines form over a range of atmospheric heights, and velocity fields are thus imprinted to varying extent on the core and wings \citep[see e.g.][]{2000A&A...359..729A,2018A&A...611A..11C}. We therefore opt to carry out a forward-modelling approach by performing line formation calculations for the solar \ion{K}{i} line. 
  Due to the pronounced departures from LTE for the \ion{K}{i} 7698\,{\AA} resonance line \citep{1992A&A...265..237B,2019A&A...627A.177R}, it is crucial to carry out full 3D non-LTE radiative transfer computations to obtain realistic atmospheric velocity information.
  
  The line formation calculations are performed using \textsc{balder} \citep{2016MNRAS.455.3735A,2016MNRAS.463.1518A,2018A&A...615A.139A}, a 3D non-LTE radiative transfer code based on the \textsc{multi3d} code \citep{1999ASSL..240..379B,2009ASPC..415...87L}.
  Our model atom contains 29 levels of \ion{K}i plus the \ion{K}{ii} ground state, and resolves the fine structure in all doublets of \ion{K}i. Atomic energy levels and oscillator strengths originate from NIST \citep{2008JPCRD..37....7S}, and collisional line broadening is computed following the method of \citet{1998PASA...15..336B}. We implement transition rates due to collisions with electrons and hydrogen atoms following \citet{2019A&A...627A.177R}. For radiative transitions our simplified atom considers only the 7664--7698\,\AA\ resonance line doublet, as their departure from LTE is almost entirely explained through photon losses in the resonance lines themselves which produces a characteristic sub-thermal source function that deepens the line \citep{2019A&A...627A.177R}. 
  This simplification was a necessary trade-off to obtain line profiles across the entire simulation time series. In any case, test calculations indicated that this two-line atom differs from a comprehensive \ion{K}{i} model atom with 134 levels and 250 bound-bound radiative transitions \citep{2019A&A...627A.177R} by about 1\,\% in the depth of the 7698\,\AA\ line and with negligible differences in inferred radial velocities.
  Moreover, we rescaled the hydrodynamic simulations from a resolution of $240^3$ to $120^2 \times 220$. We solve the statistical equilibrium by computing the monochromatic radiation field using 26 short characteristics (eight polar angles, and four azimuthal angles for each non-vertical ray), and typically find convergence after six accelerated lambda iterations. The emergent spectra are computed with 57 rays (seven outgoing polar angles, and eight azimuthal angles for each non-vertical ray), sampling the spectral line at a spectral resolution of 40\,\ms.
  
  We adopted a solar abundance $A(\text{K}) = 5.10$ as this produced good agreement with observations of the \ion{K}{i} 7698\,\AA\ line, but did not fine-tune this value. Figure~\ref{fig:KIline} demonstrates the spatially averaged, disk integrated \ion{K}{i} line profiles computed with \textsc{balder}, which are in excellent agreement with the observed line profile. The calculations were carried out for every snapshot in the 3D solar simulation (2880 in total) to obtain the temporal evolution of the solar \ion{K}{i} line. The location and shape of the line varies from snapshot to snapshot as a consequence of varying line-of-sight velocity fields in the line forming region.
  
  Radial velocities are extracted from our theoretical \ion{K}{i} lines in a way that is fully consistent with the BiSON observational setup. Under a magnetic field of 0.2\,T, which is approximately the magnetic field strength imposed in the BiSON spectrometer \citep{1978MNRAS.185....1B}, the \ion{K}{i} 7698\,\AA\ line is split into two components centred at $\lambda_- = 7698.8907$ {\AA} and $\lambda_+ = 7699.0384$ {\AA} due to the Zeeman effect. Assuming the apparatus holds a temperature of 400 K yields a thermal broadening of 400\,\ms for the two laboratory lines. Their Gaussian profiles are schematically shown in Fig.~\ref{fig:KIline}. We compute the convolution between the spatially and temporally averaged theoretical solar \ion{K}{i} line profile and each of the laboratory lines:
\begin{equation} \label{eq:IBIR}
\begin{aligned}
F_{\rm B} &= f_\lambda * G_{\lambda_-}
\\
F_{\rm R} &= f_\lambda * G_{\lambda_+},
\end{aligned}
\end{equation}
and subsequently the normalised flux difference
\begin{equation} \label{eq:fluxdiff}
\mathcal{R} = \frac{ F_{\rm B} - F_{\rm R} }{ F_{\rm B} + F_{\rm R} },
\end{equation}
where ``*'' is the convolution operator, $f_\lambda$ is the solar flux spectrum and $G_{\lambda_{-(+)}}$ is the blue (red) component of laboratory potassium line after Zeeman splitting\footnote{We note that the magnitude of $G_{\lambda_{-(+)}}$ has no effect on our results, as it cancels out in Eq.~\eqref{eq:fluxdiff}}. The normalised flux difference $\mathcal{R}$ is proportional to the radial velocity \citep{1996SoPh..168....1C}. In order to quantify the proportionality constant, we translate the averaged \ion{K}{i} line back and forth in velocity space in steps of 3\,\ms.
The aforementioned calculation (Eqs.~\eqref{eq:IBIR} and \eqref{eq:fluxdiff}) is repeated each time to obtain the velocity shift as a function of $\mathcal{R}$, which is well described by a linear function 
\begin{equation} \label{eq:v-R}
\mathfrak{v} = 3020.576 \mathcal{R} - 556.903 \; {\rm m\,s^{-1}},
\end{equation}
over the interval $[-500,500]$\,\ms. The slope of Eq.~\eqref{eq:v-R} is in good agreement with the proportionality constant used in BiSON\footnote{In practice, the diurnal change of the measured normalized flux difference $\mathcal{R}$ due to the rotation of the Earth is used to calibrate the proportionality constant. A third-order polynomial relation, calibrated on a daily basis, is used.}, which is typically 3000\,\ms\ \citep{1996SoPh..168....1C}, implying that our synthesised line profile describes the solar \ion{K}{i} line in a realistic way.

  The above procedure is identical to the means by which BiSON extracts radial velocity from the solar \ion{K}{i} line. We therefore apply Eqs.~\eqref{eq:IBIR}-\eqref{eq:v-R} to every snapshot in the 3D solar simulation. The thus evaluated radial velocity $\mathfrak{v}$ as a function of time has the same physical meaning as what BiSON measures.
The fluctuation of radial velocity is mainly caused by radial oscillations in the simulation box. We note that temporal variations of granulation also contribute to a velocity fluctuation. However, by performing a horizontal average over more than ten granules, the influence of granulation on velocity fluctuation largely cancels out (\citealt{2000A&A...359..729A}, Sect.~4.2). Finally, we apply a Fourier transform to the temporal evolution of $\mathfrak{v}$ into the frequency domain to obtain the radial velocity power spectrum presented in Fig.~\ref{fig:PSRV}. The radial velocity amplitude of the dominant simulation mode is 81.6\,\ms\ at frequency 3.299 mHz.

\subsection{Simulating the SONG radial velocity signal} \label{sec:SONG}

\begin{figure*}
\begin{overpic}[width=\textwidth]{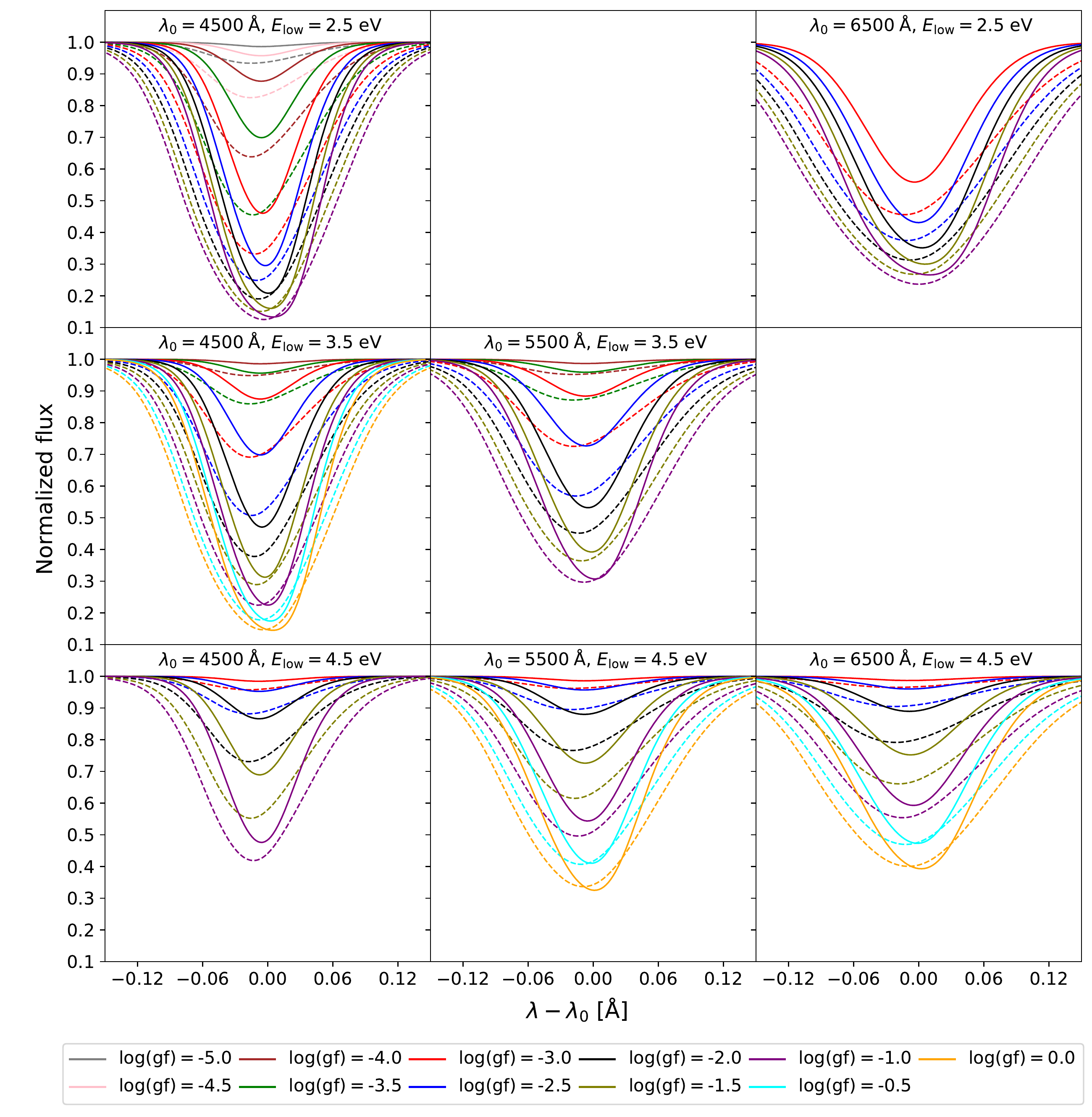}
\end{overpic}
\caption{Spatially averaged \ion{Fe}{i} line profiles of all line parameters tabulated in Table \ref{tb:line} computed from \textsc{scate}. Results from one example simulation snapshot are shown here for both the Sun (solid lines) and $\epsilon$ Tau (dashed lines). Reference wavelength $\lambda_0$ and $E_{\rm low}$ are marked in the figures, and $\log\rm (gf)$ values are colour-coded as indicated in the legend. The top middle and middle right panels are left blank, because the corresponding $\{ \lambda, E_{\rm low} \}$ combinations are not selected as representative line parameters (see Table \ref{tb:line}).
}
\label{fig:FeI}
\end{figure*}

\begin{figure}
\begin{overpic}[width=\columnwidth]{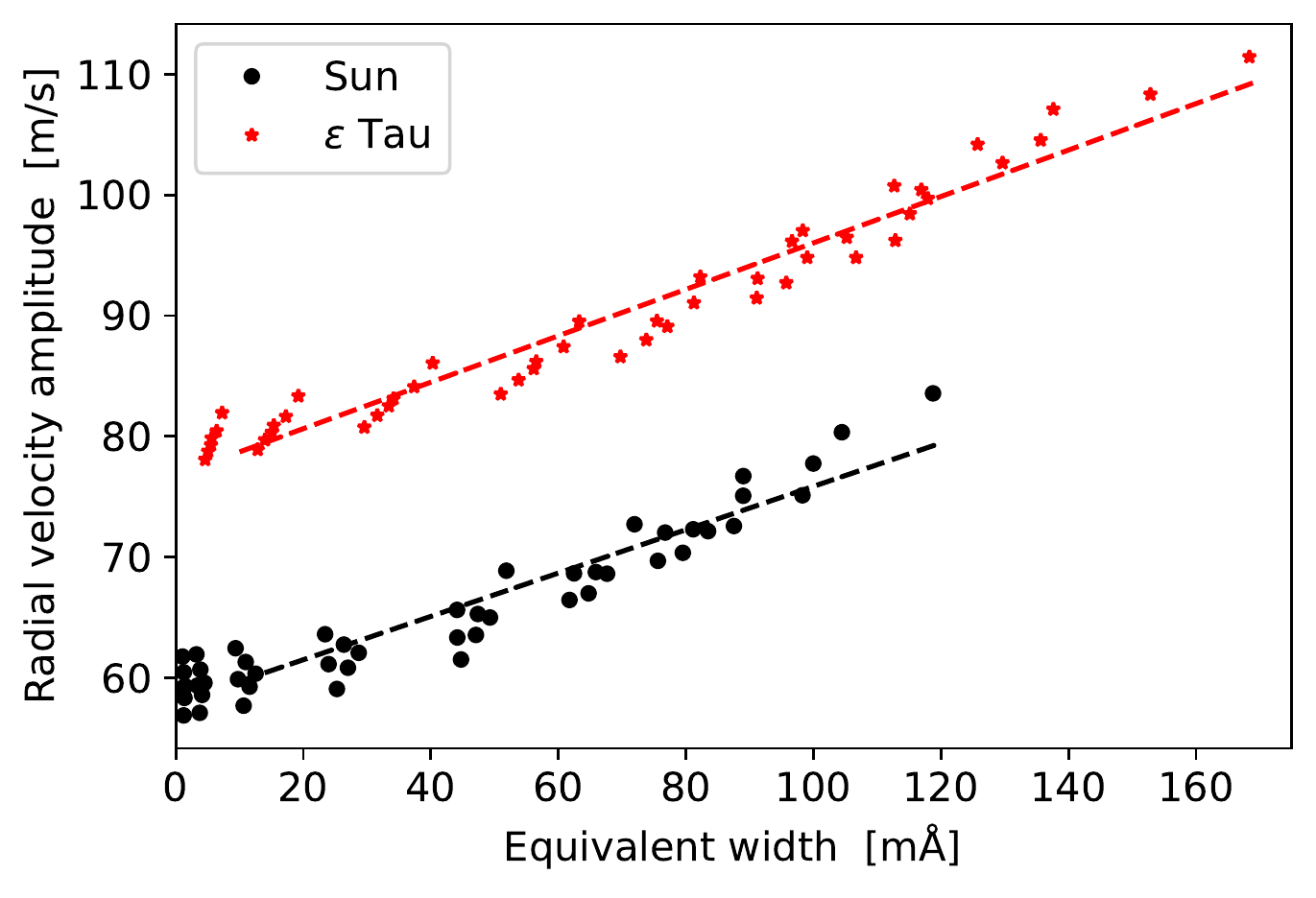}
\end{overpic}
\caption{Radial velocity amplitude at the frequency of the dominant simulation mode evaluated from 49 fictitious \ion{Fe}{i} lines are plotted against the equivalent width of these lines. Results from the solar and red giant simulations are shown in black dots and red asterisks, respectively. Linear fits to these data points are presented in dashed lines.}
\label{fig:RVEW}
\end{figure}

\begin{table}
	\centering
	\caption{Selected parameters of fictitious \ion{Fe}{i} lines. Here $\log\rm (gf)$ values are distributed in steps of 0.5. There are totally 49 lines covering typical \ion{Fe}{i} lines in the Sun and $\epsilon$ Tau within the SONG wavelength range. 
	}
	\label{tb:line}
	\begin{tabular}{ccc}
    \toprule[2pt]
 Wavelength ({\AA})           & $E_{\rm low}$ (eV)   & $\log\rm (gf)$
\\
	\midrule[1pt]
 4500 & 2.5 & [$-5$, $-1$]
\\
      & 3.5 & [$-4$, 0]
\\
      & 4.5 & [$-3$, $-1$]                     
\\
 5500 & 3.5 & [$-4$, $-1$]    
\\
      & 4.5 & [$-3$, 0]
\\
 6500 & 2.5 & [$-3$, $-1$]
\\
      & 4.5 & [$-3$, 0]
\\
	\bottomrule[2pt]
	\end{tabular}
\end{table}

  SONG is a high-resolution spectrograph that covers a broad wavelength range, 4400--6900\,\AA, containing thousands of spectral absorption lines. For computational reasons, it is not feasible to perform high-resolution 3D radiative transfer over such a broad wavelength range for our very long hydrodynamic time series. Instead, we develop a simplified method that relies on computing synthetic spectra for representative lines covering a range of atomic line properties. 

  First, in order to have a macroscopic perception of absorption lines in the wavelength region covered by SONG, we computed the strength of every atomic absorption line between 4400 and 6900 {\AA} with the TurboSpectrum code (v15.1; \citealt{2012ascl.soft05004P}) for the stellar parameters of the Sun and $\epsilon$ Tau, using an atomic linelist from the Vienna Atomic Line Database (VALD, \citealt{2015PhyS...90e4005R}).
  Selecting only lines with a line strength $W_\lambda/\lambda \ge 10^{-5}$ (corresponding to an equivalent width $W_\lambda = 5$\,m\AA\ at $\lambda = 5000$\,\AA), we find a total of 4100 atomic lines for the Sun, and 7000 for $\epsilon$ Tau. Of these, 40\,\% are due to \ion{Fe}{i}, and another 40\,\% are due to neutral species of other Fe-peak elements. These numbers are in good agreement with the list of lines identified over the same wavelength range in the spectrum of Arcturus by \citet{2000vnia.book.....H}. As the vast majority of spectral lines in the optical region are due to the neutral species of iron or elements with similar electron structure to iron, we use \ion{Fe}{i} as a representative species. 
  
  The problem now is to determine a set of \ion{Fe}{i} lines that can reasonably simulate SONG observations. The strength and shape of an absorption line is mainly governed by three parameters: the wavelength $\lambda$, which controls the background opacity due largely to H$^-$, the excitation potential of the lower ionization state $E_{\rm low}$, which determines the population of the level in LTE and thus the number of absorbers, and the oscillator strength $\log\rm(gf)$, which represents roughly the likelihood that a photon is absorbed by the line.
  It is therefore necessary to select representative values of $\lambda$, $E_{\rm low}$ and $\log\rm(gf)$ combinations, ensuring the corresponding artificial \ion{Fe}{i} lines cover the properties of most observed \ion{Fe}{i} lines. 
  Firstly, we choose three wavelength ranges: 4400--4600 {\AA}, 5400--5600 {\AA}, and 6400--6600 {\AA}. In a given wavelength region, we pick all \ion{Fe}{i} lines with equivalent width greater than 10 m{\AA} from the aforementioned solar and $\epsilon$ Tau line lists. All selected lines from the two theoretical line lists are then classified into different groups based on their $E_{\rm low}$. For example, the thus determined typical $E_{\rm low}$ values for \ion{Fe}{i} lines between 4400 and 4600 {\AA} are 2.5, 3.5 and 4.5 eV. Next, for a given $E_{\rm low}$, we further select a set of $\log\rm(gf)$ values that span the entire $\log\rm(gf)$ range seen in the theoretical line list. The selection procedure gives 49 line parameter combinations listed in Table \ref{tb:line}, which are reasonable representation of \ion{Fe}{i} lines from the Sun and $\epsilon$ Tau between 4400 and 6900 {\AA}.
  
  We carry out 3D LTE line formation calculations for these 49 fictitious \ion{Fe}i lines using the \textsc{scate} code \citep{2011A&A...529A.158H}. In these high-resolution calculations, we tabulate continuum opacities and photon destruction probabilities for a set of temperatures and densitites and use these to compute the effects of continuum scattering at run time. Line opacities are likewise evaluated at run time, taking into account local velocity fields that produce Doppler shifts in the line profiles. Specific intensities are computed for 20 rays, along five different polar and four azimuthal angles; the former are distributed on a Gauss-Legendre quadrature and the latter are equidistant. We compute the spectral lines at high resolution, with a velocity step of just 40\,\ms, over a range $\pm 10$\,k\ms\ that samples the entire line. We perform these calculations on each snapshot from the simulations of the Sun and $\epsilon$ Tau, and note that in the case of $\epsilon$ Tau, the Fe abundance used in the line formation calculation is adopted from the reference metallicity $\rm [Fe/H] = 0.15$.
  Example theoretical line profiles, i.e.~the normalized emergent flux as a function of wavelength, computed from one simulation snapshot of the Sun and $\epsilon$ Tau are shown in Fig.~\ref{fig:FeI}. The theoretical lines in $\epsilon$ Tau are broader than the corresponding solar lines for all line parameters, due to the larger velocity field in red giant stars.
  
  The line formation calculation introduced above gives the time evolution of all 49 fictitious \ion{Fe}{i} lines, from which we can then extract the corresponding radial velocity variation. As reference, we use template spectra computed as the temporal averages from the two simulation sequences. For each $\{\lambda, E_{\rm low}, \log\rm(gf) \}$ combination, a certain theoretical \ion{Fe}{i} line from a given snapshot is fitted to the template line profile using a $\chi^2$ technique to obtain the wavelength shift $\Delta\lambda$ of this line. This method closely resembles the cross correlation technique often used in observational work \citep[e.g.][]{1996PASP..108..500B}.

  The thus obtained radial velocity as a function of time is translated to the frequency domain through a Fourier transform, which gives radial velocity amplitude at the frequency of the dominant simulation mode. Radial velocity amplitudes for the 49 lines range from roughly 60 to 80\,\ms\ in the solar case and 80--110\,\ms in the case of $\epsilon$ Tau, as shown in Fig.~\ref{fig:RVEW}. The next question is how then to reliably determine a final radial velocity amplitude, given these 49 different values. Recall that strong absorption lines tend to form higher up in the stellar photosphere than weak lines \citep[e.g.][]{2003rtsa.book.....R}, where the velocity fields are typically larger due to the substantially smaller densities.
  It is therefore anticipated that the magnitude of radial velocity amplitude is correlated with the strength of the line, and Fig.~\ref{fig:RVEW} indeed follows an approximately linear relationship between the radial velocity amplitude and equivalent width for the \ion{Fe}{i} lines considered for both stars. 

  We select from our theoretical line list every \ion{Fe}i line in the SONG spectral range with equivalent width between 10 and 200\,m\AA, and use our fitted linear relations to estimate the radial velocity variation amplitude for each line. We exclude weaker lines as these are unlikely to significantly influence the radial velocity determination in a real stellar spectrum. Very strong lines with $W_{\lambda} > 200$ m{\AA} are also excluded, because extrapolations to even higher equivalent width are not guaranteed to be reliable. In total, 1407 and 2224 lines are selected this way for the Sun and $\epsilon$ Tau, respectively.
  The ensemble of estimated radial velocity amplitudes are averaged to a final value, weighted by the equivalent width. The weighted average is performed with the understanding that the signal to noise ratio of weak lines is typically smaller than stronger lines, meaning a relatively larger error and thus a smaller influence on the final result (see Fig.~2 of \citealt{2013MNRAS.435.1563A}).
  
  Although the method developed here is not identical to the means by which SONG determines radial velocity from the observed spectra, it simulates the SONG observations sufficiently well. First, the set of fictitious \ion{Fe}{i} lines carefully chosen in this work is able to represent the properties of most \ion{Fe}{i} lines seen between 4400 and 6900 {\AA}, which constitute a large part of all lines in this wavelength interval. Second, the procedure to extract radial velocity from theoretical spectral lines is similar to how radial velocities are typically obtained from observed spectra.
  Third, the evaluation of our final radial velocity amplitude includes the information of many spectral lines that span the whole range in observation. The major uncertainty in our method is associated with the linear relationship between radial velocity amplitude and equivalent width. Due to the complicated physical processes involved in spectral line formation in a 3D atmosphere \citep[e.g.][]{2000A&A...359..729A}, it is difficult to quantify higher order effects beyond the linear relation between $\mathfrak{v}$ and $W_{\lambda}$; that is, the systematic uncertainty of the linear fitting. Nevertheless, it is still illuminating to provide the statistical uncertainty. 
  The statistical uncertainty is quantified using the bootstrap method. The data set considered here is the radial velocity amplitude and equivalent width of 49 fictitious \ion{Fe}{i} lines. We conduct 10000 bootstrap samplings, that is, generating 10000 data sets each containing 49 randomly sampled $\mathfrak{v}$ and $W_{\lambda}$ pairs.
  A linear regression between equivalent width and radial velocity is then performed for each re-sampled data set. For each fitting, we compute the equivalent width weighted mean radial velocity amplitude for all selected \ion{Fe}{i} lines. The bootstrap method therefore results in 10000 weighted mean radial velocity amplitudes, their mean and variance is the desired final radial velocity amplitude and its statistical uncertainty, which is $72.2 \pm 0.5$\,\ms\ for the 3D solar model and $93.2 \pm 0.3$\,\ms\ for the $\epsilon$ Tau model.

\section{Results}

\begin{table}
\begin{threeparttable}
\centering
\caption{Summary of predicted and observed oscillation amplitudes and amplitude ratios for the Sun and $\epsilon$ Tau. Here we emphasize again that individual oscillation amplitudes from 3D atmosphere simulations are not comparable to the corresponding observations (cf.~Sect.~\ref{sec:3Dmodel} and \ref{sec:Proof}).
\label{tb:result}
}
{\begin{tabular*}{\columnwidth}{@{\extracolsep{\fill}}ccc}
\toprule[2pt]
  Sun & Modelling & Observation
  \\
\midrule[1pt]
  Bolometric (ppm) & $1714.0 \pm 3.0$           & $3.58 \pm 0.16$ (a)
  \\ 
  BiSON (\ms)      & 81.6           & $0.187 \pm 0.007$ (b)
  \\ 
  SONG (\ms)       & $72.2 \pm 0.5$ & $0.166 \pm 0.004$ (c)
  \\
  Bolometric/BiSON (ppm/[\ms]) & $21.0 \pm 0.04$ & $19.1 \pm 1.1$
  \\
  Bolometric/SONG (ppm/[\ms])  & $23.7 \pm 0.2$  & $21.6 \pm 1.1$
  \\
\midrule[1pt]
  $\epsilon$ Tau & Modelling & Observation
  \\
\midrule[1pt]
  Bolometric (ppm)    & $3070.3 \pm 7.4$  & ---
  \\
  $\textit{K2}$ (ppm) & $4515.1 \pm 10.9$ & $39.8 \pm 1.4$ (d)
  \\
  SONG (\ms)          & $93.2 \pm 0.3$    & $0.94 \pm 0.04$ (d)
  \\
  \textit{K2}/SONG (ppm/[\ms]) & $48.4 \pm 0.2$ & $42.2 \pm 2.3$ (d)
  \\
\bottomrule[2pt]
\end{tabular*}}

    \begin{tablenotes}
      \item Reference: (a): \cite{2009A&A...495..979M}; 
      (b): \cite{2008ApJ...682.1370K}; 
      (c): \cite{2019A&A...623L...9F};
      (d): \cite{2019A&A...622A.190A}
    \end{tablenotes}
    
\end{threeparttable}
\end{table}

  Our results, together with the corresponding observations, are summarised in Table \ref{tb:result}.
  
  The predicted ratio between luminosity and BiSON radial velocity amplitude at approximately 3.3 mHz is 1714.0 ppm $\div$ 81.6\,\ms\ $\approx$ 21.0 ppm/[\ms]. Observationally, the measured maximum bolometric amplitude per radial mode for the Sun is $3.58 \pm 0.16$ ppm according to \citet{2009A&A...495..979M}; the value presented in their paper is multiplied by $\sqrt{2}$ to convert the root mean square luminosity variation to luminosity amplitude. The peak radial velocity amplitude per radial mode measured by BiSON is $18.7 \pm 0.7$\,c\ms\ \citep{2008ApJ...682.1370K}. Therefore, the amplitude ratio determined from observation is approximately $19.1 \pm 1.1$ ppm/[\ms], in good agreement with the result predicted by our simulations. However, we caution that the above observed solar amplitude ratio is evaluated at the frequency of maximum power $\nu_{\max}$, that is, 3.1 mHz for the Sun \citep{2008ApJ...682.1370K}, whereas our theoretical result is obtained at the frequency of the dominant simulation mode (3.3 mHz). The two values are hence not strictly comparable because amplitude ratio depends on frequency in principle. Nevertheless, the frequency dependence is weak, especially for frequencies near $\nu_{\max}$, as shown in detailed asteroseismic observations (for example, Fig.~13 of \citealt{2019A&A...622A.190A}). Therefore, as an initial effort to this topic, a single amplitude ratio value is likely to be sufficient to describe the relationship between luminosity and radial velocity variation for a given star.  
  
  The predicted ratio between luminosity and SONG radial velocity amplitudes (at approximately 3.3 mHz) is $23.7 \pm 0.2$ ppm/[\ms] for the Sun. We emphasise that the presented uncertainty reflects the combined statistical error of the granulation background fitting (Sect.~\ref{sec:Abol}) and the linear fit to radial velocities (Sect.~\ref{sec:SONG}).
  On the other hand, the measured solar maximum bolometric amplitude is $3.58 \pm 0.16$ ppm, whereas the maximum radial velocity amplitude obtained from SONG observations of the Sun is $16.6 \pm 0.4$\,c\ms\ \citep{2019A&A...623L...9F}. Together, the observed solar amplitude ratio is $21.6 \pm 1.1$ ppm/[\ms], being consistent with our theoretical result. 
  For $\epsilon$ Tau, the amplitude ratio measured by \citet{2019A&A...622A.190A} is $42.2 \pm 2.3$ ppm/[\ms], which is the ratio between \textit{K2} luminosity amplitude and the SONG radial velocity amplitude. The intrinsic luminosity amplitude computed from our simulations is $3070.3 \pm 7.4$ ppm at the frequency of the dominant simulation mode, and the conversion factor between intrinsic and \textit{Kepler} luminosity amplitude is 0.68 for $\epsilon$ Tau as quantified in Sect.~\ref{sec:cP-bol}. According to Eq.~\eqref{eq:Abol-AP}, the theoretical luminosity amplitude in the \textit{Kepler} band-pass turns out to be $4515.1 \pm 10.9$ ppm. Dividing this value by the theoretical SONG radial velocity amplitude gives the predicted amplitude ratio between \textit{Kepler} (\textit{K2}) and SONG for $\epsilon$ Tau, which is $48.4 \pm 0.2$ ppm/[\ms]. Again, we find reasonable agreement between our theoretical calculations and the observations.
  
  We note that the amplitude ratio estimated from the widely used empirical relationship of \citet[their Eq.~5]{1995A&A...293...87K} is 23.2 ppm/[\ms] for $\epsilon$ Tau, being significantly lower than the observed value by almost a factor of two. The good agreement between observation and our theoretical result based on detailed modelling therefore shows great potential to accurately quantify the relationship between luminosity and radial velocity amplitude, especially for red giant stars where the empirical amplitude ratio relation may fail. Nonetheless, we are aware that our predicted ratios are systematically larger than corresponding observations by about 10\%. The underlying reason for this small discrepancy is not entirely clear and will be investigated in future work.

\section{Conclusions}

  In this work, we investigated the relationship between photometric and spectroscopic measurements of solar-like oscillations using  3D radiative-hydrodynamical stellar atmosphere simulations with the \textsc{Stagger} code. We used as test cases the Sun and the Hyades red giant $\epsilon$ Tau.
  Our simulations provide realistic descriptions of fluid motions from first principles, hence naturally yield compressible effects such as sound waves. Although sound waves emerging in the simulation domain are analogous to $p$-modes in solar-type oscillating stars, the simulation modes have much larger oscillation amplitudes than observed stellar $p$-modes due to the limited extent of simulation box. Therefore, we first analytically demonstrated that 3D simulations are still able to reliably predict the ratio between luminosity and velocity amplitudes, despite the individual amplitude values not being comparable with observations.
  
  Having established the basis of our analysis, we computed the spectrum of luminosity variation based on bolometric fluxes predicted by the state-of-the-art radiative transfer module in our simulation. Contribution from the granulation background was modelled in a way similar to what is applied to real observations. The modelled granulation background was then subtracted from the luminosity spectrum to obtain the intrinsic luminosity amplitude of the dominant simulation mode. To enable comparison with amplitudes measured with a given spacecraft, it was necessary to quantify the conversion factor (also called bolometric correction) between the intrinsic and measured luminosity amplitudes. We adopted the theoretical formulation of \citet{2009A&A...495..979M}, \citet{2011A&A...531A.124B} and \citet{2019MNRAS.489.1072L} for the evaluation of the conversion factor, whose components are consistently computed via 3D spectrum synthesis using the code \textsc{scate}. As an initial step, we evaluated the conversion factor between intrinsic and \textit{Kepler} luminosity amplitude for the two stars studied in this work. For $\epsilon$ Tau, our result differs from \citet{2019MNRAS.489.1072L} by roughly 10\%, implying that the conversion factors of red giant stars are likely to be sensitive to the choice of model atmosphere.
  
  In turn, we have developed novel methods to simulate the spectroscopic measurement of stellar oscillations from numerical simulations for the first time. Theoretical radial velocities are obtained from realistic spectral line formation calculations with 3D time-dependent atmosphere models as input. 
  In order to simulate BiSON, which measures solar oscillations through the \ion{K}{i} line, we performed detailed 3D non-LTE \ion{K}{i} line formation calculations with \textsc{balder} for our solar model atmosphere. The computed \ion{K}{i} line profile is in excellent agreement with observation, its temporal evolution (that is, Doppler shift) gives radial velocities whose physical meaning is identical to what measured by BiSON.
  In addition, we carried out 3D LTE line formation calculations for a large set of fictitious \ion{Fe}{i} lines using \textsc{scate} to simulate SONG observations which determines radial velocities from a forest of absorption lines between 4400 and 6900 {\AA}. The parameters of the chosen \ion{Fe}{i} lines were carefully selected such that their properties cover most lines typically seen in the Sun and a warm giant within the SONG wavelength range. For each selected line, radial velocities were extracted according to the Doppler shift of line profiles with method that resemble observations. This procedure was repeated for all selected lines, thereby giving rise to a set of independent radial velocity amplitudes. With the insight that the radial velocity amplitude computed from a certain line is correlated with its strength, we fit a linear function between radial velocity amplitude and equivalent width based on results from all selected fictitious lines. The linear relation was further used to estimate radial velocity amplitudes for all visible lines within the SONG wavelength range in our theoretical line list, which were subsequently reduced to a final radial velocity amplitude value via weighted average. 
  
  In concert, the calculations gave us the ratio between luminosity and radial velocity amplitude, which characterize the relationship between photometric and spectroscopic measurement of stellar oscillations. Given the 3D atmosphere simulations and line formation calculations presented in this work, our approach to quantify the amplitude ratio is free from any empirical parameters that have to be assumed or calibrated from observations.
  The \textit{ab initio} nature of our numerical modelling therefore not only reveals the underlying physics behind asteroseismic observations but also enables an independent comparison between theoretical results and observed amplitude ratio.
  For the Sun, our theoretical bolometric and BiSON ratio as well as bolometric and SONG ratio are compared with helioseismic observations with good agreements, thus validate our numerical approach. In the case of $\epsilon$ Tau, the predicted ratio between \textit{K2} and SONG amplitude matches corresponding observations as well, which is particularly encouraging as the observed amplitude ratio of this star cannot be explained by the widely used empirical amplitude ratio scaling relation. 
  The good theoretical--observational consistency achieved for both the Sun and a red giant star suggested that our method of connecting the luminosity and radial velocity measurements of solar-like oscillations is robust, effective, and likely applicable to a wide range of stellar parameters. This demonstrates great potential in the era of simultaneous observations of stellar oscillation with both space-photometry (such as TESS) and ground-based spectroscopy (such as SONG).
  
  In the future we plan to extend our analysis to cover the parameter space of solar-like oscillating stars across the HR diagram (i.e.~dwarfs, subgiants and red giants), which will give amplitude ratios as a function of basic stellar parameters. These theoretical amplitude ratios can provide valuable insight to asteroseismic observations by helping to determine whether a star is better observed in photometry or spectroscopy. Conversely, it is possible to determine radial velocity oscillation amplitude from \textit{Kepler} or TESS data through the theoretical amplitude ratio relation, thereby quantifying the oscillation part of the so-called ``radial velocity jitter'' (see \citealt{2018MNRAS.480L..48Y} for a pioneering study in this direction) which is of great importance in exoplanet science.

\section*{Acknowledgements}

  The authors are grateful to Remo Collet, Tim Bedding and Saskia Hekker for valuable comments and fruitful discussions. This project has been supported by the Australian Research Council (project DP150100250 awarded to MA and LC). LC is the recipient of an ARC Future Fellowship (project FT160100402). MJ was also supported by the Research School of Astronomy and Astrophysics at the Australian National University and funding from Australian Research Council grant No.\ DP150100250. AMA acknowledges support from the Swedish Research Council (VR 2016-03765), and the project grant `The New Milky Way' (KAW 2013.0052) from the Knut and Alice Wallenberg Foundation. MA gratefully acknowledges additional funding through an ARC Laureate Fellowship (project FL110100012). This work was supported by computational resources provided by the Australian Government through the National Computational Infrastructure (NCI) facility under the ANU Merit Allocation Scheme and the National Computational Merit Allocation Scheme. Parts of this research were conducted by the Australian Research Council Centre of Excellence for All Sky Astrophysics in 3 Dimensions (ASTRO 3D), through project number CE170100013.

\section*{Data availability}

  Data available on request. The data underlying this article will be shared on reasonable request to the corresponding author.




\bibliographystyle{mnras}
\bibliography{References} 







\bsp	
\label{lastpage}
\end{document}